\hsize=31pc
\vsize=49pc
\lineskip=0pt
\parskip=0pt plus 1pt
\hfuzz=1pt
\vfuzz=2pt
\pretolerance=2500
\tolerance=5000
\vbadness=5000
\hbadness=5000
\widowpenalty=500
\clubpenalty=200
\brokenpenalty=500
\predisplaypenalty=200
\voffset=-1pc
\nopagenumbers
\catcode`@=11
\newif\ifams
\amsfalse 
%
%
%
\newfam\bdifam
\newfam\bsyfam
\newfam\bssfam
\newfam\msafam
\newfam\msbfam
\newif\ifxxpt
\newif\ifxviipt
\newif\ifxivpt
\newif\ifxiipt
\newif\ifxipt
\newif\ifxpt
\newif\ifixpt
\newif\ifviiipt
\newif\ifviipt
\newif\ifvipt
\newif\ifvpt
%
%
\def\headsize#1#2{\def\headb@seline{#2}%
                \ifnum#1=20\def\HEAD{twenty}%
                           \def\smHEAD{twelve}%
                           \def\vsHEAD{nine}%
                           \ifxxpt\else\xdef\f@ntsize{\HEAD}%
                           \def\m@g{4}\def\s@ze{20.74}%
                           \loadheadfonts\xxpttrue\fi
                           \ifxiipt\else\xdef\f@ntsize{\smHEAD}%
                           \def\m@g{1}\def\s@ze{12}%
                           \loadxiiptfonts\xiipttrue\fi
                           \ifixpt\else\xdef\f@ntsize{\vsHEAD}%
                           \def\s@ze{9}%
                           \loadsmallfonts\ixpttrue\fi
                      \else
                \ifnum#1=17\def\HEAD{seventeen}%
                           \def\smHEAD{eleven}%
                           \def\vsHEAD{eight}%
                           \ifxviipt\else\xdef\f@ntsize{\HEAD}%
                           \def\m@g{3}\def\s@ze{17.28}%
                           \loadheadfonts\xviipttrue\fi
                           \ifxipt\else\xdef\f@ntsize{\smHEAD}%
                           \loadxiptfonts\xipttrue\fi
                           \ifviiipt\else\xdef\f@ntsize{\vsHEAD}%
                           \def\s@ze{8}%
                           \loadsmallfonts\viiipttrue\fi
                      \else\def\HEAD{fourteen}%
                           \def\smHEAD{ten}%
                           \def\vsHEAD{seven}%
                           \ifxivpt\else\xdef\f@ntsize{\HEAD}%
                           \def\m@g{2}\def\s@ze{14.4}%
                           \loadheadfonts\xivpttrue\fi
                           \ifxpt\else\xdef\f@ntsize{\smHEAD}%
                           \def\s@ze{10}%
                           \loadxptfonts\xpttrue\fi
                           \ifviipt\else\xdef\f@ntsize{\vsHEAD}%
                           \def\s@ze{7}%
                           \loadviiptfonts\viipttrue\fi
                \ifnum#1=14\else
                \message{Header size should be 20, 17 or 14 point
                              will now default to 14pt}\fi
                \fi\fi\headfonts}
%
%
\def\textsize#1#2{\def\textb@seline{#2}%
                 \ifnum#1=12\def\TEXT{twelve}%
                           \def\smTEXT{eight}%
                           \def\vsTEXT{six}%
                           \ifxiipt\else\xdef\f@ntsize{\TEXT}%
                           \def\m@g{1}\def\s@ze{12}%
                           \loadxiiptfonts\xiipttrue\fi
                           \ifviiipt\else\xdef\f@ntsize{\smTEXT}%
                           \def\s@ze{8}%
                           \loadsmallfonts\viiipttrue\fi
                           \ifvipt\else\xdef\f@ntsize{\vsTEXT}%
                           \def\s@ze{6}%
                           \loadviptfonts\vipttrue\fi
                      \else
                \ifnum#1=11\def\TEXT{eleven}%
                           \def\smTEXT{seven}%
                           \def\vsTEXT{five}%
                           \ifxipt\else\xdef\f@ntsize{\TEXT}%
                           \def\s@ze{11}%
                           \loadxiptfonts\xipttrue\fi
                           \ifviipt\else\xdef\f@ntsize{\smTEXT}%
                           \loadviiptfonts\viipttrue\fi
                           \ifvpt\else\xdef\f@ntsize{\vsTEXT}%
                           \def\s@ze{5}%
                           \loadvptfonts\vpttrue\fi
                      \else\def\TEXT{ten}%
                           \def\smTEXT{seven}%
                           \def\vsTEXT{five}%
                           \ifxpt\else\xdef\f@ntsize{\TEXT}%
                           \loadxptfonts\xpttrue\fi
                           \ifviipt\else\xdef\f@ntsize{\smTEXT}%
                           \def\s@ze{7}%
                           \loadviiptfonts\viipttrue\fi
                           \ifvpt\else\xdef\f@ntsize{\vsTEXT}%
                           \def\s@ze{5}%
                           \loadvptfonts\vpttrue\fi
                \ifnum#1=10\else
                \message{Text size should be 12, 11 or 10 point
                              will now default to 10pt}\fi
                \fi\fi\textfonts}
%
%
\def\smallsize#1#2{\def\smallb@seline{#2}%
                 \ifnum#1=10\def\SMALL{ten}%
                           \def\smSMALL{seven}%
                           \def\vsSMALL{five}%
                           \ifxpt\else\xdef\f@ntsize{\SMALL}%
                           \loadxptfonts\xpttrue\fi
                           \ifviipt\else\xdef\f@ntsize{\smSMALL}%
                           \def\s@ze{7}%
                           \loadviiptfonts\viipttrue\fi
                           \ifvpt\else\xdef\f@ntsize{\vsSMALL}%
                           \def\s@ze{5}%
                           \loadvptfonts\vpttrue\fi
                       \else
                 \ifnum#1=9\def\SMALL{nine}%
                           \def\smSMALL{six}%
                           \def\vsSMALL{five}%
                           \ifixpt\else\xdef\f@ntsize{\SMALL}%
                           \def\s@ze{9}%
                           \loadsmallfonts\ixpttrue\fi
                           \ifvipt\else\xdef\f@ntsize{\smSMALL}%
                           \def\s@ze{6}%
                           \loadviptfonts\vipttrue\fi
                           \ifvpt\else\xdef\f@ntsize{\vsSMALL}%
                           \def\s@ze{5}%
                           \loadvptfonts\vpttrue\fi
                       \else
                           \def\SMALL{eight}%
                           \def\smSMALL{six}%
                           \def\vsSMALL{five}%
                           \ifviiipt\else\xdef\f@ntsize{\SMALL}%
                           \def\s@ze{8}%
                           \loadsmallfonts\viiipttrue\fi
                           \ifvipt\else\xdef\f@ntsize{\smSMALL}%
                           \def\s@ze{6}%
                           \loadviptfonts\vipttrue\fi
                           \ifvpt\else\xdef\f@ntsize{\vsSMALL}%
                           \def\s@ze{5}%
                           \loadvptfonts\vpttrue\fi
                 \ifnum#1=8\else\message{Small size should be 10, 9 or
                            8 point will now default to 8pt}\fi
                \fi\fi\smallfonts}
\def\F@nt{\expandafter\font\csname}
\def\Sk@w{\expandafter\skewchar\csname}
\def\@nd{\endcsname}
\def\@step#1{ scaled \magstep#1}
\def\@half{ scaled \magstephalf}
\def\@t#1{ at #1pt}
%
%
\def\loadheadfonts{\bigf@nts
\F@nt \f@ntsize bdi\@nd=cmmib10 \@t{\s@ze}%
\Sk@w \f@ntsize bdi\@nd='177
\F@nt \f@ntsize bsy\@nd=cmbsy10 \@t{\s@ze}%
\Sk@w \f@ntsize bsy\@nd='60
\F@nt \f@ntsize bss\@nd=cmssbx10 \@t{\s@ze}}
%
%
\def\loadxiiptfonts{\bigf@nts
\F@nt \f@ntsize bdi\@nd=cmmib10 \@step{\m@g}%
\Sk@w \f@ntsize bdi\@nd='177
\F@nt \f@ntsize bsy\@nd=cmbsy10 \@step{\m@g}%
\Sk@w \f@ntsize bsy\@nd='60
\F@nt \f@ntsize bss\@nd=cmssbx10 \@step{\m@g}}
%
%
\def\loadxiptfonts{%
\font\elevenrm=cmr10 \@half
\font\eleveni=cmmi10 \@half
\skewchar\eleveni='177
\font\elevensy=cmsy10 \@half
\skewchar\elevensy='60
\font\elevenex=cmex10 \@half
\font\elevenit=cmti10 \@half
\font\elevensl=cmsl10 \@half
\font\elevenbf=cmbx10 \@half
\font\eleventt=cmtt10 \@half
\ifams\font\elevenmsa=msam10 \@half
\font\elevenmsb=msbm10 \@half\else\fi
\font\elevenbdi=cmmib10 \@half
\skewchar\elevenbdi='177
\font\elevenbsy=cmbsy10 \@half
\skewchar\elevenbsy='60
\font\elevenbss=cmssbx10 \@half}
%
%
\def\loadxptfonts{%
\font\tenbdi=cmmib10
\skewchar\tenbdi='177
\font\tenbsy=cmbsy10
\skewchar\tenbsy='60
\ifams\font\tenmsa=msam10
\font\tenmsb=msbm10\else\fi
\font\tenbss=cmssbx10}%
%
%
\def\loadsmallfonts{\smallf@nts
\ifams
\F@nt \f@ntsize ex\@nd=cmex\s@ze
\else
\F@nt \f@ntsize ex\@nd=cmex10\fi
\F@nt \f@ntsize it\@nd=cmti\s@ze
\F@nt \f@ntsize sl\@nd=cmsl\s@ze
\F@nt \f@ntsize tt\@nd=cmtt\s@ze}
%
%
\def\loadviiptfonts{%
\font\sevenit=cmti7
\font\sevensl=cmsl8 at 7pt
\ifams\font\sevenmsa=msam7
\font\sevenmsb=msbm7
\font\sevenex=cmex7
\font\sevenbsy=cmbsy7
\font\sevenbdi=cmmib7\else
\font\sevenex=cmex10
\font\sevenbsy=cmbsy10 at 7pt
\font\sevenbdi=cmmib10 at 7pt\fi
\skewchar\sevenbsy='60
\skewchar\sevenbdi='177
\font\sevenbss=cmssbx10 at 7pt}%
%
%
\def\loadviptfonts{\smallf@nts
\ifams\font\sixex=cmex7 at 6pt\else
\font\sixex=cmex10\fi
\font\sixit=cmti7 at 6pt}
%
%
\def\loadvptfonts{%
\font\fiveit=cmti7 at 5pt
\ifams\font\fiveex=cmex7 at 5pt
\font\fivebdi=cmmib5
\font\fivebsy=cmbsy5
\font\fivemsa=msam5
\font\fivemsb=msbm5\else
\font\fiveex=cmex10
\font\fivebdi=cmmib10 at 5pt
\font\fivebsy=cmbsy10 at 5pt\fi
\skewchar\fivebdi='177
\skewchar\fivebsy='60
\font\fivebss=cmssbx10 at 5pt}
\def\bigf@nts{%
\F@nt \f@ntsize rm\@nd=cmr10 \@step{\m@g}%
\F@nt \f@ntsize i\@nd=cmmi10 \@step{\m@g}%
\Sk@w \f@ntsize i\@nd='177
\F@nt \f@ntsize sy\@nd=cmsy10 \@step{\m@g}%
\Sk@w \f@ntsize sy\@nd='60
\F@nt \f@ntsize ex\@nd=cmex10 \@step{\m@g}%
\F@nt \f@ntsize it\@nd=cmti10 \@step{\m@g}%
\F@nt \f@ntsize sl\@nd=cmsl10 \@step{\m@g}%
\F@nt \f@ntsize bf\@nd=cmbx10 \@step{\m@g}%
\F@nt \f@ntsize tt\@nd=cmtt10 \@step{\m@g}%
\ifams
\F@nt \f@ntsize msa\@nd=msam10 \@step{\m@g}%
\F@nt \f@ntsize msb\@nd=msbm10 \@step{\m@g}\else\fi}
\def\smallf@nts{%
\F@nt \f@ntsize rm\@nd=cmr\s@ze
\F@nt \f@ntsize i\@nd=cmmi\s@ze
\Sk@w \f@ntsize i\@nd='177
\F@nt \f@ntsize sy\@nd=cmsy\s@ze
\Sk@w \f@ntsize sy\@nd='60
\F@nt \f@ntsize bf\@nd=cmbx\s@ze
\ifams
\F@nt \f@ntsize bdi\@nd=cmmib\s@ze
\F@nt \f@ntsize bsy\@nd=cmbsy\s@ze
\F@nt \f@ntsize msa\@nd=msam\s@ze
\F@nt \f@ntsize msb\@nd=msbm\s@ze
\else
\F@nt \f@ntsize bdi\@nd=cmmib10 \@t{\s@ze}%
\F@nt \f@ntsize bsy\@nd=cmbsy10 \@t{\s@ze}\fi
\Sk@w \f@ntsize bdi\@nd='177
\Sk@w \f@ntsize bsy\@nd='60
\F@nt \f@ntsize bss\@nd=cmssbx10 \@t{\s@ze}}%
%
%
\def\headfonts{%
\textfont0=\csname\HEAD rm\@nd
\scriptfont0=\csname\smHEAD rm\@nd
\scriptscriptfont0=\csname\vsHEAD rm\@nd
\def\rm{\fam0\csname\HEAD rm\@nd
\def\sc{\csname\smHEAD rm\@nd}}%
\textfont1=\csname\HEAD i\@nd
\scriptfont1=\csname\smHEAD i\@nd
\scriptscriptfont1=\csname\vsHEAD i\@nd
\textfont2=\csname\HEAD sy\@nd
\scriptfont2=\csname\smHEAD sy\@nd
\scriptscriptfont2=\csname\vsHEAD sy\@nd
\textfont3=\csname\HEAD ex\@nd
\scriptfont3=\csname\smHEAD ex\@nd
\scriptscriptfont3=\csname\smHEAD ex\@nd
\textfont\itfam=\csname\HEAD it\@nd
\scriptfont\itfam=\csname\smHEAD it\@nd
\scriptscriptfont\itfam=\csname\vsHEAD it\@nd
\def\it{\fam\itfam\csname\HEAD it\@nd
\def\sc{\csname\smHEAD it\@nd}}%
\textfont\slfam=\csname\HEAD sl\@nd
\def\sl{\fam\slfam\csname\HEAD sl\@nd
\def\sc{\csname\smHEAD sl\@nd}}%
\textfont\bffam=\csname\HEAD bf\@nd
\scriptfont\bffam=\csname\smHEAD bf\@nd
\scriptscriptfont\bffam=\csname\vsHEAD bf\@nd
\def\bf{\fam\bffam\csname\HEAD bf\@nd
\def\sc{\csname\smHEAD bf\@nd}}%
\textfont\ttfam=\csname\HEAD tt\@nd
\def\tt{\fam\ttfam\csname\HEAD tt\@nd}%
\textfont\bdifam=\csname\HEAD bdi\@nd
\scriptfont\bdifam=\csname\smHEAD bdi\@nd
\scriptscriptfont\bdifam=\csname\vsHEAD bdi\@nd
\def\bdi{\fam\bdifam\csname\HEAD bdi\@nd}%
\textfont\bsyfam=\csname\HEAD bsy\@nd
\scriptfont\bsyfam=\csname\smHEAD bsy\@nd
\def\bsy{\fam\bsyfam\csname\HEAD bsy\@nd}%
\textfont\bssfam=\csname\HEAD bss\@nd
\scriptfont\bssfam=\csname\smHEAD bss\@nd
\scriptscriptfont\bssfam=\csname\vsHEAD bss\@nd
\def\bss{\fam\bssfam\csname\HEAD bss\@nd}%
\ifams
\textfont\msafam=\csname\HEAD msa\@nd
\scriptfont\msafam=\csname\smHEAD msa\@nd
\scriptscriptfont\msafam=\csname\vsHEAD msa\@nd
\textfont\msbfam=\csname\HEAD msb\@nd
\scriptfont\msbfam=\csname\smHEAD msb\@nd
\scriptscriptfont\msbfam=\csname\vsHEAD msb\@nd
\else\fi
\normalbaselineskip=\headb@seline pt%
\setbox\strutbox=\hbox{\vrule height.7\normalbaselineskip
depth.3\baselineskip width0pt}%
\def\sc{\csname\smHEAD rm\@nd}\normalbaselines\bf}
%
%
\def\textfonts{%
\textfont0=\csname\TEXT rm\@nd
\scriptfont0=\csname\smTEXT rm\@nd
\scriptscriptfont0=\csname\vsTEXT rm\@nd
\def\rm{\fam0\csname\TEXT rm\@nd
\def\sc{\csname\smTEXT rm\@nd}}%
\textfont1=\csname\TEXT i\@nd
\scriptfont1=\csname\smTEXT i\@nd
\scriptscriptfont1=\csname\vsTEXT i\@nd
\textfont2=\csname\TEXT sy\@nd
\scriptfont2=\csname\smTEXT sy\@nd
\scriptscriptfont2=\csname\vsTEXT sy\@nd
\textfont3=\csname\TEXT ex\@nd
\scriptfont3=\csname\smTEXT ex\@nd
\scriptscriptfont3=\csname\smTEXT ex\@nd
\textfont\itfam=\csname\TEXT it\@nd
\scriptfont\itfam=\csname\smTEXT it\@nd
\scriptscriptfont\itfam=\csname\vsTEXT it\@nd
\def\it{\fam\itfam\csname\TEXT it\@nd
\def\sc{\csname\smTEXT it\@nd}}%
\textfont\slfam=\csname\TEXT sl\@nd
\def\sl{\fam\slfam\csname\TEXT sl\@nd
\def\sc{\csname\smTEXT sl\@nd}}%
\textfont\bffam=\csname\TEXT bf\@nd
\scriptfont\bffam=\csname\smTEXT bf\@nd
\scriptscriptfont\bffam=\csname\vsTEXT bf\@nd
\def\bf{\fam\bffam\csname\TEXT bf\@nd
\def\sc{\csname\smTEXT bf\@nd}}%
\textfont\ttfam=\csname\TEXT tt\@nd
\def\tt{\fam\ttfam\csname\TEXT tt\@nd}%
\textfont\bdifam=\csname\TEXT bdi\@nd
\scriptfont\bdifam=\csname\smTEXT bdi\@nd
\scriptscriptfont\bdifam=\csname\vsTEXT bdi\@nd
\def\bdi{\fam\bdifam\csname\TEXT bdi\@nd}%
\textfont\bsyfam=\csname\TEXT bsy\@nd
\scriptfont\bsyfam=\csname\smTEXT bsy\@nd
\def\bsy{\fam\bsyfam\csname\TEXT bsy\@nd}%
\textfont\bssfam=\csname\TEXT bss\@nd
\scriptfont\bssfam=\csname\smTEXT bss\@nd
\scriptscriptfont\bssfam=\csname\vsTEXT bss\@nd
\def\bss{\fam\bssfam\csname\TEXT bss\@nd}%
\ifams
\textfont\msafam=\csname\TEXT msa\@nd
\scriptfont\msafam=\csname\smTEXT msa\@nd
\scriptscriptfont\msafam=\csname\vsTEXT msa\@nd
\textfont\msbfam=\csname\TEXT msb\@nd
\scriptfont\msbfam=\csname\smTEXT msb\@nd
\scriptscriptfont\msbfam=\csname\vsTEXT msb\@nd
\else\fi
\normalbaselineskip=\textb@seline pt
\setbox\strutbox=\hbox{\vrule height.7\normalbaselineskip
depth.3\baselineskip width0pt}%
\everymath{}%
\def\sc{\csname\smTEXT rm\@nd}\normalbaselines\rm}
%
%
\def\smallfonts{%
\textfont0=\csname\SMALL rm\@nd
\scriptfont0=\csname\smSMALL rm\@nd
\scriptscriptfont0=\csname\vsSMALL rm\@nd
\def\rm{\fam0\csname\SMALL rm\@nd
\def\sc{\csname\smSMALL rm\@nd}}%
\textfont1=\csname\SMALL i\@nd
\scriptfont1=\csname\smSMALL i\@nd
\scriptscriptfont1=\csname\vsSMALL i\@nd
\textfont2=\csname\SMALL sy\@nd
\scriptfont2=\csname\smSMALL sy\@nd
\scriptscriptfont2=\csname\vsSMALL sy\@nd
\textfont3=\csname\SMALL ex\@nd
\scriptfont3=\csname\smSMALL ex\@nd
\scriptscriptfont3=\csname\smSMALL ex\@nd
\textfont\itfam=\csname\SMALL it\@nd
\scriptfont\itfam=\csname\smSMALL it\@nd
\scriptscriptfont\itfam=\csname\vsSMALL it\@nd
\def\it{\fam\itfam\csname\SMALL it\@nd
\def\sc{\csname\smSMALL it\@nd}}%
\textfont\slfam=\csname\SMALL sl\@nd
\def\sl{\fam\slfam\csname\SMALL sl\@nd
\def\sc{\csname\smSMALL sl\@nd}}%
\textfont\bffam=\csname\SMALL bf\@nd
\scriptfont\bffam=\csname\smSMALL bf\@nd
\scriptscriptfont\bffam=\csname\vsSMALL bf\@nd
\def\bf{\fam\bffam\csname\SMALL bf\@nd
\def\sc{\csname\smSMALL bf\@nd}}%
\textfont\ttfam=\csname\SMALL tt\@nd
\def\tt{\fam\ttfam\csname\SMALL tt\@nd}%
\textfont\bdifam=\csname\SMALL bdi\@nd
\scriptfont\bdifam=\csname\smSMALL bdi\@nd
\scriptscriptfont\bdifam=\csname\vsSMALL bdi\@nd
\def\bdi{\fam\bdifam\csname\SMALL bdi\@nd}%
\textfont\bsyfam=\csname\SMALL bsy\@nd
\scriptfont\bsyfam=\csname\smSMALL bsy\@nd
\def\bsy{\fam\bsyfam\csname\SMALL bsy\@nd}%
\textfont\bssfam=\csname\SMALL bss\@nd
\scriptfont\bssfam=\csname\smSMALL bss\@nd
\scriptscriptfont\bssfam=\csname\vsSMALL bss\@nd
\def\bss{\fam\bssfam\csname\SMALL bss\@nd}%
\ifams
\textfont\msafam=\csname\SMALL msa\@nd
\scriptfont\msafam=\csname\smSMALL msa\@nd
\scriptscriptfont\msafam=\csname\vsSMALL msa\@nd
\textfont\msbfam=\csname\SMALL msb\@nd
\scriptfont\msbfam=\csname\smSMALL msb\@nd
\scriptscriptfont\msbfam=\csname\vsSMALL msb\@nd
\else\fi
\normalbaselineskip=\smallb@seline pt%
\setbox\strutbox=\hbox{\vrule height.7\normalbaselineskip
depth.3\baselineskip width0pt}%
\everymath{}%
\def\sc{\csname\smSMALL rm\@nd}\normalbaselines\rm}%
\everydisplay{\indenteddisplay
   \gdef\labeltype{\eqlabel}}%
%
%
\def\hexnumber@#1{\ifcase#1 0\or 1\or 2\or 3\or 4\or 5\or 6\or 7\or 8\or
 9\or A\or B\or C\or D\or E\or F\fi}
\edef\bffam@{\hexnumber@\bffam}
\edef\bdifam@{\hexnumber@\bdifam}
\edef\bsyfam@{\hexnumber@\bsyfam}
\def\undefine#1{\let#1\undefined}
\def\newsymbol#1#2#3#4#5{\let\next@\relax
 \ifnum#2=\thr@@\let\next@\bdifam@\else
 \ifams
 \ifnum#2=\@ne\let\next@\msafam@\else
 \ifnum#2=\tw@\let\next@\msbfam@\fi\fi
 \fi\fi
 \mathchardef#1="#3\next@#4#5}
\def\mathhexbox@#1#2#3{\relax
 \ifmmode\mathpalette{}{\m@th\mathchar"#1#2#3}%
 \else\leavevmode\hbox{$\m@th\mathchar"#1#2#3$}\fi}

\def\bi#1{{\fam\bdifam\relax#1}}
%
%
\ifams\input amsmacro\fi
%
%
\newsymbol\bitGamma 3000
\newsymbol\bitDelta 3001
\newsymbol\bitTheta 3002
\newsymbol\bitLambda 3003
\newsymbol\bitXi 3004
\newsymbol\bitPi 3005
\newsymbol\bitSigma 3006
\newsymbol\bitUpsilon 3007
\newsymbol\bitPhi 3008
\newsymbol\bitPsi 3009
\newsymbol\bitOmega 300A
\newsymbol\balpha 300B
\newsymbol\bbeta 300C
\newsymbol\bgamma 300D
\newsymbol\bdelta 300E
\newsymbol\bepsilon 300F
\newsymbol\bzeta 3010
\newsymbol\bfeta 3011
\newsymbol\btheta 3012
\newsymbol\biota 3013
\newsymbol\bkappa 3014
\newsymbol\blambda 3015
\newsymbol\bmu 3016
\newsymbol\bnu 3017
\newsymbol\bxi 3018
\newsymbol\bpi 3019
\newsymbol\brho 301A
\newsymbol\bsigma 301B
\newsymbol\btau 301C
\newsymbol\bupsilon 301D
\newsymbol\bphi 301E
\newsymbol\bchi 301F
\newsymbol\bpsi 3020
\newsymbol\bomega 3021
\newsymbol\bvarepsilon 3022
\newsymbol\bvartheta 3023
\newsymbol\bvaromega 3024
\newsymbol\bvarrho 3025
\newsymbol\bvarzeta 3026
\newsymbol\bvarphi 3027
\newsymbol\bpartial 3040
\newsymbol\bell 3060
\newsymbol\bimath 307B
\newsymbol\bjmath 307C
\mathchardef\binfty "0\bsyfam@31
\mathchardef\bnabla "0\bsyfam@72
\mathchardef\bdot "2\bsyfam@01
\mathchardef\bGamma "0\bffam@00
\mathchardef\bDelta "0\bffam@01
\mathchardef\bTheta "0\bffam@02
\mathchardef\bLambda "0\bffam@03
\mathchardef\bXi "0\bffam@04
\mathchardef\bPi "0\bffam@05
\mathchardef\bSigma "0\bffam@06
\mathchardef\bUpsilon "0\bffam@07
\mathchardef\bPhi "0\bffam@08
\mathchardef\bPsi "0\bffam@09
\mathchardef\bOmega "0\bffam@0A
\mathchardef\itGamma "0100
\mathchardef\itDelta "0101
\mathchardef\itTheta "0102
\mathchardef\itLambda "0103
\mathchardef\itXi "0104
\mathchardef\itPi "0105
\mathchardef\itSigma "0106
\mathchardef\itUpsilon "0107
\mathchardef\itPhi "0108
\mathchardef\itPsi "0109
\mathchardef\itOmega "010A
\mathchardef\Gamma "0000
\mathchardef\Delta "0001
\mathchardef\Theta "0002
\mathchardef\Lambda "0003
\mathchardef\Xi "0004
\mathchardef\Pi "0005
\mathchardef\Sigma "0006
\mathchardef\Upsilon "0007
\mathchardef\Phi "0008
\mathchardef\Psi "0009
\mathchardef\Omega "000A
%
%
\newcount\firstpage  \firstpage=1  
\newcount\jnl                      
\newcount\secno                    
\newcount\subno                    
\newcount\subsubno                 
\newcount\appno                    
\newcount\tabno                    
\newcount\figno                    
\newcount\countno                  
\newcount\refno                    
\newcount\eqlett     \eqlett=97    
\newif\ifletter
\newif\ifwide
\newif\ifnotfull
\newif\ifaligned
\newif\ifnumbysec
\newif\ifappendix
\newif\ifnumapp
\newif\ifssf
\newif\ifppt
\newdimen\t@bwidth
\newdimen\c@pwidth
\newdimen\digitwidth                    
\newdimen\argwidth                      
\newdimen\secindent    \secindent=5pc   
\newdimen\textind    \textind=16pt      
\newdimen\tempval                       
\newskip\beforesecskip
\def\beforesecspace{\vskip\beforesecskip\relax}
\newskip\beforesubskip
\def\beforesubspace{\vskip\beforesubskip\relax}
\newskip\beforesubsubskip
\def\beforesubsubspace{\vskip\beforesubsubskip\relax}
\newskip\secskip
\def\secspace{\vskip\secskip\relax}
\newskip\subskip
\def\subspace{\vskip\subskip\relax}
\newskip\insertskip
\def\insertspace{\vskip\insertskip\relax}
\def\sp@ce{\ifx\next*\let\next=\@ssf
               \else\let\next=\@nossf\fi\next}
\def\@ssf#1{\nobreak\secspace\global\ssftrue\nobreak}
\def\@nossf{\nobreak\secspace\nobreak\noindent\ignorespaces}
\def\subsp@ce{\ifx\next*\let\next=\@sssf
               \else\let\next=\@nosssf\fi\next}
\def\@sssf#1{\nobreak\subspace\global\ssftrue\nobreak}
\def\@nosssf{\nobreak\subspace\nobreak\noindent\ignorespaces}
\beforesecskip=24pt plus12pt minus8pt
\beforesubskip=12pt plus6pt minus4pt
\beforesubsubskip=12pt plus6pt minus4pt
\secskip=12pt plus 2pt minus 2pt
\subskip=6pt plus3pt minus2pt
\insertskip=18pt plus6pt minus6pt%
\fontdimen16\tensy=2.7pt
\fontdimen17\tensy=2.7pt
%
%
\def\eqlabel{(\ifappendix\applett
               \ifnumbysec\ifnum\secno>0 \the\secno\fi.\fi
               \else\ifnumbysec\the\secno.\fi\fi\the\countno)}
\def\seclabel{\ifappendix\ifnumapp\else\applett\fi
    \ifnum\secno>0 \the\secno
    \ifnumbysec\ifnum\subno>0.\the\subno\fi\fi\fi
    \else\the\secno\fi\ifnum\subno>0.\the\subno
         \ifnum\subsubno>0.\the\subsubno\fi\fi}
\def\tablabel{\ifappendix\applett\fi\the\tabno}
\def\figlabel{\ifappendix\applett\fi\the\figno}
\def\gac{\global\advance\countno by 1}
%
%

\def\vfootnote#1{\insert\footins\bgroup
\interlinepenalty=\interfootnotelinepenalty
\splittopskip=\ht\strutbox 
\splitmaxdepth=\dp\strutbox \floatingpenalty=20000
\leftskip=0pt \rightskip=0pt \spaceskip=0pt \xspaceskip=0pt%
\noindent\smallfonts\rm #1\ \ignorespaces\footstrut\futurelet\next\fo@t}
%
%
\def\endinsert{\egroup
    \if@mid \dimen@=\ht0 \advance\dimen@ by\dp0
       \advance\dimen@ by12\p@ \advance\dimen@ by\pagetotal
       \ifdim\dimen@>\pagegoal \@midfalse\p@gefalse\fi\fi
    \if@mid \insertspace \box0 \par \ifdim\lastskip<\insertskip
    \removelastskip \penalty-200 \insertspace \fi
    \else\insert\topins{\penalty100
       \splittopskip=0pt \splitmaxdepth=\maxdimen
       \floatingpenalty=0
       \ifp@ge \dimen@=\dp0
       \vbox to\vsize{\unvbox0 \kern-\dimen@}%
       \else\box0\nobreak\insertspace\fi}\fi\endgroup}
%
%
%
\def\ind{\hbox to \secindent{\hfill}}
%
%

%
%

%
%
\def\indeqn#1{\alignedfalse\displ@y\halign{\hbox to \displaywidth
    {$\ind\@lign\displaystyle##\hfil$}\crcr #1\crcr}}
%
%
\def\indalign#1{\alignedtrue\displ@y \tabskip=0pt
  \halign to\displaywidth{\ind$\@lign\displaystyle{##}$\tabskip=0pt
    &$\@lign\displaystyle{{}##}$\hfill\tabskip=\centering
    &\llap{$\@lign\hbox{\rm##}$}\tabskip=0pt\crcr
    #1\crcr}}
\def\fl{{\hskip-\secindent}}
\def\indenteddisplay#1$${\indispl@y{#1 }}
\def\indispl@y#1{\disptest#1\eqalignno\eqalignno\disptest}
\def\disptest#1\eqalignno#2\eqalignno#3\disptest{%
    \ifx#3\eqalignno
    \indalign#2%
    \else\indeqn{#1}\fi$$}
%
%

%
%

%
%

%
%

%
%

\def\ns{\noalign{\vskip-3pt}}

%

%
%
\def\bhbar{\rlap{\kern1pt\raise.4ex\hbox{\bf\char'40}}\bi{h}}
\def\case#1#2{{\textstyle{#1\over#2}}}

\def\d{{\rm d}}
\def\e{{\rm e}}
\def\etal{{\it et al\/}\ }
\def\frac#1#2{{#1\over#2}}
\ifams
\def\lap{\lesssim}
\def\gap{\gtrsim}

\let\geq=\geqslant
\else

\def\gap{\;\lower3pt\hbox{$\buildrel > \over \sim$}\;}%
\def\lap{\;\lower3pt\hbox{$\buildrel < \over \sim$}\;}\fi
\def\i{{\rm i}}
\chardef\ii="10
\def\tqs{\hbox to 25pt{\hfil}}

\def\Bbbone{1\kern-.22em {\rm l}}
%
%
\def\rp{\raise8pt\hbox{$\scriptstyle\prime$}}
%
%
%
%

%
%
\def\[#1\]{\setbox0=\hbox{$\dsty#1$}\argwidth=\wd0
    \setbox0=\hbox{$\left[\box0\right]$}\advance\argwidth by -\wd0
    \left[\kern.3\argwidth\box0\kern.3\argwidth\right]}
%
%
\def\lsb#1\rsb{\setbox0=\hbox{$#1$}\argwidth=\wd0
    \setbox0=\hbox{$\left[\box0\right]$}\advance\argwidth by -\wd0
    \left[\kern.3\argwidth\box0\kern.3\argwidth\right]}
%

%
%

%
\def\pt(#1){({\it #1\/})}
\let\dsty=\displaystyle

%
%
\def\reactions#1{\vskip 12pt plus2pt minus2pt%
\vbox{\hbox{\kern\secindent\vrule\kern12pt%
\vbox{\kern0.5pt\vbox{\hsize=24pc\parindent=0pt\smallfonts\rm NUCLEAR
REACTIONS\strut\quad #1\strut}\kern0.5pt}\kern12pt\vrule}}}
%
%
\def\slashchar#1{\setbox0=\hbox{$#1$}\dimen0=\wd0%
\setbox1=\hbox{/}\dimen1=\wd1%
\ifdim\dimen0>\dimen1%
\rlap{\hbox to \dimen0{\hfil/\hfil}}#1\else
\rlap{\hbox to \dimen1{\hfil$#1$\hfil}}/\fi}
%
%
\def\textindent#1{\noindent\hbox to \parindent{#1\hss}\ignorespaces}
%
%
\def\opencirc{\raise1pt\hbox{$\scriptstyle{\bigcirc}$}}

\ifams
\def\opensqr{\hbox{$\square$}}

\def\opentridown{\hbox{$\triangledown$}}

\else
\def\opensqr{\vbox{\hrule height.4pt\hbox{\vrule width.4pt height3.5pt
    \kern3.5pt\vrule width.4pt}\hrule height.4pt}}

\def\opentridown{\raise1pt\hbox{$\scriptstyle\bigtriangledown$}}

\fi

%
%
\def\m@th{\mathsurround=0pt}
%
%
\def\cases#1{%
\left\{\,\vcenter{\normalbaselines\openup1\jot\m@th%
     \ialign{$\displaystyle##\hfil$&\rm\tqs##\hfil\crcr#1\crcr}}\right.}%
%
%
\def\oldcases#1{\left\{\,\vcenter{\normalbaselines\m@th
    \ialign{$##\hfil$&\rm\quad##\hfil\crcr#1\crcr}}\right.}
%
%
\def\numcases#1{\left\{\,\vcenter{\baselineskip=15pt\m@th%
     \ialign{$\displaystyle##\hfil$&\rm\tqs##\hfil
     \crcr#1\crcr}}\right.\hfill
     \vcenter{\baselineskip=15pt\m@th%
     \ialign{\rlap{$\phantom{\displaystyle##\hfil}$}\tabskip=0pt&\en
     \rlap{\phantom{##\hfil}}\crcr#1\crcr}}}
\def\ptnumcases#1{\left\{\,\vcenter{\baselineskip=15pt\m@th%
     \ialign{$\displaystyle##\hfil$&\rm\tqs##\hfil
     \crcr#1\crcr}}\right.\hfill
     \vcenter{\baselineskip=15pt\m@th%
     \ialign{\rlap{$\phantom{\displaystyle##\hfil}$}\tabskip=0pt&\enpt
     \rlap{\phantom{##\hfil}}\crcr#1\crcr}}\global\eqlett=97
     \global\advance\countno by 1}
%
%
\def\eq(#1){\ifaligned\@mp(#1)\else\hfill\llap{{\rm (#1)}}\fi}
\def\ceq(#1){\ns\ns\ifaligned\@mp\fi\eq(#1)\cr\ns\ns}
\def\eqpt(#1#2){\ifaligned\@mp(#1{\it #2\/})
                    \else\hfill\llap{{\rm (#1{\it #2\/})}}\fi}

%
%
\countno=1
\def\eqnobysec{\numbysectrue}
\def\aleq{&\rm(\ifappendix\applett
               \ifnumbysec\ifnum\secno>0 \the\secno\fi.\fi
               \else\ifnumbysec\the\secno.\fi\fi\the\countno}
\def\noaleq{\hfill\llap\bgroup\rm(\ifappendix\applett
               \ifnumbysec\ifnum\secno>0 \the\secno\fi.\fi
               \else\ifnumbysec\the\secno.\fi\fi\the\countno}
\def\@mp{&}
\def\en{\ifaligned\aleq)\else\noaleq)\egroup\fi\gac}
\def\cen{\ns\ns\ifaligned\@mp\fi\en\cr\ns\ns}
\def\enpt{\ifaligned\aleq{\it\char\the\eqlett})\else
    \noaleq{\it\char\the\eqlett})\egroup\fi
    \global\advance\eqlett by 1}
\def\endpt{\ifaligned\aleq{\it\char\the\eqlett})\else
    \noaleq{\it\char\the\eqlett})\egroup\fi
    \global\eqlett=97\gac}
%
%

\def\JPA{{\it J. Phys. A: Math. Gen.}}
\def\JPC{{\it J. Phys. C: Solid State Phys.}}     



%
%

\def\APNY{{\it Ann. Phys., NY\/}}

\def\NP{{\it Nucl. Phys.}}

\def\PR{{\it Phys. Rev.}}
\def\PRL{{\it Phys. Rev. Lett.}}

\def\ZP{{\it Z. Phys.}}
\headline={\ifodd\pageno{\ifnum\pageno=\firstpage\hfill
   \else\rrhead\fi}\else\lrhead\fi}
\def\rrhead{\textfonts\hskip\secindent\it
    \shorttitle\hfill\rm\folio}
\def\lrhead{\textfonts\hbox to\secindent{\rm\folio\hss}%
    \it\aunames\hss}
\footline={\ifnum\pageno=\firstpage \hfill\textfonts\rm\folio\fi}
\def\@rticle#1#2{\vglue.5pc
    {\parindent=\secindent \bf #1\par}
     \vskip2.5pc
    {\exhyphenpenalty=10000\hyphenpenalty=10000
     \baselineskip=18pt\raggedright\noindent
     \headfonts\bf#2\par}\futurelet\next\sh@rttitle}%
\def\title#1{\gdef\shorttitle{#1}
    \vglue4pc{\exhyphenpenalty=10000\hyphenpenalty=10000
    \baselineskip=18pt
    \raggedright\parindent=0pt
    \headfonts\bf#1\par}\futurelet\next\sh@rttitle}

\def\article#1#2{\gdef\shorttitle{#2}\@rticle{#1}{#2}}
\def\review#1{\gdef\shorttitle{#1}%
    \@rticle{REVIEW \ifpbm\else ARTICLE\fi}{#1}}
\def\topical#1{\gdef\shorttitle{#1}%
    \@rticle{TOPICAL REVIEW}{#1}}
\def\comment#1{\gdef\shorttitle{#1}%
    \@rticle{COMMENT}{#1}}
\def\note#1{\gdef\shorttitle{#1}%
    \@rticle{NOTE}{#1}}
\def\prelim#1{\gdef\shorttitle{#1}%
    \@rticle{PRELIMINARY COMMUNICATION}{#1}}
\def\letter#1{\gdef\shorttitle{Letter to the Editor}%
     \gdef\aunames{Letter to the Editor}
     \global\lettertrue\ifnum\jnl=7\global\letterfalse\fi
     \@rticle{LETTER TO THE EDITOR}{#1}}
\def\sh@rttitle{\ifx\next[\let\next=\sh@rt
                \else\let\next=\f@ll\fi\next}
\def\sh@rt[#1]{\gdef\shorttitle{#1}}
\def\f@ll{}
\def\author#1{\ifletter\else\gdef\aunames{#1}\fi\vskip1.5pc
    {\parindent=\secindent
     \hang\textfonts
     \ifppt\bf\else\rm\fi#1\par}
     \ifppt\bigskip\else\smallskip\fi
     \futurelet\next\@unames}
\def\@unames{\ifx\next[\let\next=\short@uthor
                 \else\let\next=\@uthor\fi\next}
\def\short@uthor[#1]{\gdef\aunames{#1}}
\def\@uthor{}
\def\address#1{{\parindent=\secindent
    \exhyphenpenalty=10000\hyphenpenalty=10000
\ifppt\textfonts\else\smallfonts\fi\hang\raggedright\rm#1\par}%
    \ifppt\bigskip\fi}
\def\jl#1{\global\jnl=#1}
\jl{0}%
\def\journal{\ifnum\jnl=1 J. Phys.\ A: Math.\ Gen.\
        \else\ifnum\jnl=2 J. Phys.\ B: At.\ Mol.\ Opt.\ Phys.\
        \else\ifnum\jnl=3 J. Phys.:\ Condens. Matter\
        \else\ifnum\jnl=4 J. Phys.\ G: Nucl.\ Part.\ Phys.\
        \else\ifnum\jnl=5 Inverse Problems\
        \else\ifnum\jnl=6 Class. Quantum Grav.\
        \else\ifnum\jnl=7 Network\
        \else\ifnum\jnl=8 Nonlinearity\
        \else\ifnum\jnl=9 Quantum Opt.\
        \else\ifnum\jnl=10 Waves in Random Media\
        \else\ifnum\jnl=11 Pure Appl. Opt.\
        \else\ifnum\jnl=12 Phys. Med. Biol.\
        \else\ifnum\jnl=13 Modelling Simulation Mater.\ Sci.\ Eng.\
        \else\ifnum\jnl=14 Plasma Phys. Control. Fusion\
        \else\ifnum\jnl=15 Physiol. Meas.\
        \else\ifnum\jnl=16 Sov.\ Lightwave Commun.\
        \else\ifnum\jnl=17 J. Phys.\ D: Appl.\ Phys.\
        \else\ifnum\jnl=18 Supercond.\ Sci.\ Technol.\
        \else\ifnum\jnl=19 Semicond.\ Sci.\ Technol.\
        \else\ifnum\jnl=20 Nanotechnology\
        \else\ifnum\jnl=21 Meas.\ Sci.\ Technol.\
        \else\ifnum\jnl=22 Plasma Sources Sci.\ Technol.\
        \else\ifnum\jnl=23 Smart Mater.\ Struct.\
        \else\ifnum\jnl=24 J.\ Micromech.\ Microeng.\
   \else Institute of Physics Publishing\
   \fi\fi\fi\fi\fi\fi\fi\fi\fi\fi\fi\fi\fi\fi\fi
   \fi\fi\fi\fi\fi\fi\fi\fi\fi}
\let\abs=\beginabstract

\let\endabs=\endabstract
\def\submitted{\ifppt\noindent\textfonts\rm Submitted to \journal\par
     \bigskip\fi}
\def\today{\number\day\ \ifcase\month\or
     January\or February\or March\or April\or May\or June\or
     July\or August\or September\or October\or November\or
     December\fi\space \number\year}
\def\date{\ifppt\noindent\textfonts\rm
     Date: \today\par\goodbreak\bigskip\fi}
%
%
\def\pacs#1{\ifppt\noindent\textfonts\rm
     PACS number(s): #1\par\bigskip\fi}
%

%
%
\def\section#1{\ifppt\ifnum\secno=0\eject\fi\fi
    \subno=0\subsubno=0\global\advance\secno by 1
    \gdef\labeltype{\seclabel}\ifnumbysec\countno=1\fi
    \goodbreak\beforesecspace\nobreak
    \noindent{\bf \the\secno. #1}\par\futurelet\next\sp@ce}
\def\subsection#1{\subsubno=0\global\advance\subno by 1
     \gdef\labeltype{\seclabel}%
     \ifssf\else\goodbreak\beforesubspace\fi
     \global\ssffalse\nobreak
     \noindent{\it \the\secno.\the\subno. #1\par}%
     \futurelet\next\subsp@ce}
\def\subsubsection#1{\global\advance\subsubno by 1
     \gdef\labeltype{\seclabel}%
     \ifssf\else\goodbreak\beforesubsubspace\fi
     \global\ssffalse\nobreak
     \noindent{\it \the\secno.\the\subno.\the\subsubno. #1}\null.
     \ignorespaces}
%

%
%
\def\numappendix#1{\ifappendix\ifnumbysec\countno=1\fi\else
    \countno=1\figno=0\tabno=0\fi
    \subno=0\global\advance\appno by 1
    \secno=\appno\gdef\applett{A}\gdef\labeltype{\seclabel}%
    \global\appendixtrue\global\numapptrue
    \goodbreak\beforesecspace\nobreak
    \noindent{\bf Appendix \the\appno. #1\par}%
    \futurelet\next\sp@ce}
\def\numsubappendix#1{\global\advance\subno by 1\subsubno=0
    \gdef\labeltype{\seclabel}%
    \ifssf\else\goodbreak\beforesubspace\fi
    \global\ssffalse\nobreak
    \noindent{\it A\the\appno.\the\subno. #1\par}%
    \futurelet\next\subsp@ce}
\def\@ppendix#1#2#3{\countno=1\subno=0\subsubno=0\secno=0\figno=0\tabno=0
    \gdef\applett{#1}\gdef\labeltype{\seclabel}\global\appendixtrue
    \goodbreak\beforesecspace\nobreak
    \noindent{\bf Appendix#2#3\par}\futurelet\next\sp@ce}
\def\Appendix#1{\@ppendix{A}{. }{#1}}
\def\appendix#1#2{\@ppendix{#1}{ #1. }{#2}}
\def\App#1{\@ppendix{A}{ }{#1}}
\def\app{\@ppendix{A}{}{}}
\def\subappendix#1#2{\global\advance\subno by 1\subsubno=0
    \gdef\labeltype{\seclabel}%
    \ifssf\else\goodbreak\beforesubspace\fi
    \global\ssffalse\nobreak
    \noindent{\it #1\the\subno. #2\par}%
    \nobreak\subspace\noindent\ignorespaces}
%
%
\def\@ck#1{\ifletter\bigskip\noindent\ignorespaces\else
    \goodbreak\beforesecspace\nobreak
    \noindent{\bf Acknowledgment#1\par}%
    \nobreak\secspace\noindent\ignorespaces\fi}
\def\ack{\@ck{s}}
\def\ackn{\@ck{}}
\def\n@ip#1{\goodbreak\beforesecspace\nobreak
    \noindent\smallfonts{\it #1}. \rm\ignorespaces}
\def\naip{\n@ip{Note added in proof}}
\def\na{\n@ip{Note added}}

%
%

%

%
%

%

%
\def\table#1{\tablecaption{#1}}
\def\tablecont{\topinsert\global\advance\tabno by -1
    \tablecaption{(continued)}}
\def\tablecaption#1{\gdef\labeltype{\tablabel}\global\widefalse
    \leftskip=\secindent\parindent=0pt
    \global\advance\tabno by 1
    \smallfonts{\bf Table \ifappendix\applett\fi\the\tabno.} \rm #1\par
    \smallskip\futurelet\next\t@b}
\def\endtable{\vfill\goodbreak}
\def\t@b{\ifx\next*\let\next=\widet@b
             \else\ifx\next[\let\next=\fullwidet@b
                      \else\let\next=\narrowt@b\fi\fi
             \next}
\def\widet@b#1{\global\widetrue\global\notfulltrue
    \t@bwidth=\hsize\advance\t@bwidth by -\secindent}
\def\fullwidet@b[#1]{\global\widetrue\global\notfullfalse
    \leftskip=0pt\t@bwidth=\hsize}
\def\narrowt@b{\global\notfulltrue}
\def\align{\catcode`?=13\ifnotfull\moveright\secindent\fi
    \vbox\bgroup\halign\ifwide to \t@bwidth\fi
    \bgroup\strut\tabskip=1.2pc plus1pc minus.5pc}
\def\endalign{\egroup\egroup\catcode`?=12}

%
%
\def\L#1{#1\hfill}

%
%
\def\br{\noalign{\vskip2pt\hrule height1pt\vskip2pt}}

%

%
%
\def\centre#1#2{\multispan{#1}{\hfill#2\hfill}}
\def\crule#1{\multispan{#1}{\hrulefill}}

\catcode`?=13
\def\lineup{\setbox0=\hbox{\smallfonts\rm 0}%
    \digitwidth=\wd0%
    \def?{\kern\digitwidth}%
    \def\\{\hbox{$\phantom{-}$}}%
    \def\-{\llap{$-$}}}
\catcode`?=12
%
%
\def\sidetable#1#2{\hbox{\ifppt\hsize=18pc\t@bwidth=18pc
                          \else\hsize=15pc\t@bwidth=15pc\fi
    \parindent=0pt\vtop{\null #1\par}%
    \ifppt\hskip1.2pc\else\hskip1pc\fi
    \vtop{\null #2\par}}}
\def\lstable#1#2{\everypar{}\tempval=\hsize\hsize=\vsize
    \vsize=\tempval\hoffset=-3pc
    \global\tabno=#1\gdef\labeltype{\tablabel}%
    \noindent\smallfonts{\bf Table \ifappendix\applett\fi
    \the\tabno.} \rm #2\par
    \smallskip\futurelet\next\t@b}
\def\inctabno{\global\advance\tabno by 1}
%
%

%

%
\def\figure#1{\figc@ption{#1}\bigskip}
\def\figc@ption#1{\global\advance\figno by 1\gdef\labeltype{\figlabel}%
   {\parindent=\secindent\smallfonts\hang
    {\bf Figure \ifappendix\applett\fi\the\figno.} \rm #1\par}}
%
%
\def\refHEAD{\goodbreak\beforesecspace
     \noindent\textfonts{\bf References}\par
     \let\ref=\rf
     \nobreak\smallfonts\rm}
\def\numreferences{\refHEAD\parindent=30pt
     \everypar{\hang\noindent\frenchspacing\rm}
     \secspace}
\def\rf#1{\par\noindent\hbox to 21pt{\hss #1\quad}\ignorespaces}
%

%

%
%

%
%

%
%

%
%

%
\catcode`\@=12
%
%

%
%
\def\jnlstyle{\pptfalse\headsize{14}{18}%
\textsize{10}{12}%
\smallsize{8}{10}
\textind=16pt}
%
%

%
%

%
\parindent=\textind
%
%
\catcode`@=11
\newwrite\auxfile
\newwrite\xreffile
\newif\ifxrefwarning \xrefwarningtrue
\newif\ifauxfile
\newif\ifxreffile
\def\testforxref{\begingroup
    \immediate\openin\xreffile = \jobname.xrf\space
    \ifeof\xreffile\global\xreffilefalse
    \else\global\xreffiletrue\fi
    \immediate\closein\xreffile
    \endgroup}
\def\testforaux{\begingroup
    \immediate\openin\auxfile = \jobname.aux\space
    \ifeof\auxfile\global\auxfilefalse
    \else\global\auxfiletrue\fi
    \immediate\closein\auxfile
    \endgroup}
\def\openreffile{\immediate\openout\auxfile = \jobname.aux}%
\def\readreffile{%
    \testforxref
    \testforaux
    \ifxreffile
       \begingroup
         \@setletters
         \input \jobname.xrf
       \endgroup
    \else
\message{No cross-reference file existed, some labels may be undefined}%
    \fi\openreffile}%
\def\@setletters{%
    \catcode`_=11 \catcode`+=11
    \catcode`-=11 \catcode`@=11
    \catcode`0=11 \catcode`1=11
    \catcode`2=11 \catcode`3=11
    \catcode`4=11 \catcode`5=11
    \catcode`6=11 \catcode`7=11
    \catcode`8=11 \catcode`9=11
    \catcode`(=11 \catcode`)=11
    \catcode`:=11 \catcode`'=11
    \catcode`&=11 \catcode`;=11
    \catcode`.=11}%
\gdef\el@b{\eqlabel}
\gdef\sl@b{\seclabel}
\gdef\tl@b{\tablabel}
\gdef\fl@b{\figlabel}
\def\l@belno{\ifx\labeltype\el@b
   \let\labelno=\en\def\@label{\eqlabel}%
   \else\let\labelno=\ignorespaces
   \ifx\labeltype\sl@b \def\@label{\seclabel}%
   \else\ifx\labeltype\tl@b \def\@label{\tablabel}%
   \else\ifx\labeltype\fl@b \def\@label{\figlabel}%
   \else\def\@label{\seclabel}\fi\fi\fi\fi}
\def\label#1{\l@belno\expandafter\xdef\csname #1@\endcsname{\@label}%
    \immediate\write\auxfile{\string
    \gdef\expandafter\string\csname @#1\endcsname{\@label}}%
    \labelno}%
\def\ref#1{%
    \expandafter \ifx \csname @#1\endcsname\relax
    \message{Undefined label `#1'.}%
    \expandafter\xdef\csname @#1\endcsname{(??)}\fi
    \csname @#1\endcsname}%
\readreffile
%
%
\def\bibitem#1{\global\advance\refno by 1%
    \immediate\write\auxfile{\string
    \gdef\expandafter\string\csname #1@\endcsname{\the\refno}}%
    \rf{[\the\refno]}}%
\def\bitem[#1]#2{\immediate\write\auxfile{\string
    \gdef\expandafter\string\csname #2@\endcsname{#1}}%
    \rf{[#1]}}%
\def\cite#1{\hbox{[\splitarg{#1}]}}%
\def\splitarg#1{\@pt#1,\@ptend}%
\def\@pt#1,#2\@ptend{\ifempty{#1}\else
    \@pttwo #1\@pttwoend
    \ifempty{#2}\else\sp@cer\@pt#2\@ptend\fi\fi}%
\def\@pttwo#1\@pttwoend{\expandafter
    \ifx \csname#1@\endcsname\@pttwoend\else
    \@ifundefined{#1}{{\bf ?}%
    \message{Undefined citation `#1' on page
    \the\pageno}}{\csname#1@\endcsname}\fi}%
\def\@pttwoend{@@@@@}%
\def\sp@cer{,\nobreak\thinspace}%
\def\ifempty#1{\@ifempty #1\@xx\@xxx}%
\def\@ifempty#1#2\@xxx{\ifx #1\@xx}%
\def\@xx{@@@@}%
\def\@xxx{@@@@}%
\long\def\@ifundefined#1#2#3{\expandafter\ifx\csname
  #1@\endcsname\relax#2\else#3\fi}%
\catcode`@=12

\magnification1200
\input epsf
\eqnobysec

\def\appendix{\goodbreak\beforesecspace
     \noindent\textfonts{\bf Appendix}\secspace}
\def\figure#1{\global\advance\figno by 1\gdef\labeltype{\figlabel}%
   {\parindent=\secindent\smallfonts\hang
    {\bf Figure \ifappendix\applett\fi\the\figno.} \rm #1\par}}
\def\lpsn#1#2{LPSN-#1-LT#2}
\def\endtable{\parindent=\textind\textfonts\rm\bigskip}
\headline={\ifodd\pageno{\ifnum\pageno=\firstpage\titlehead
   \else\rrhead\fi}\else\lrhead\fi}
\footline={\ifnum\pageno=\firstpage{\smallfonts cond-mat/9505111}
\hfil\textfonts\rm\folio\fi}
\def\titlehead{\smallfonts To appear in J. Phys. A: Math. Gen.
\hfil\lpsn{95}{2}}

\firstpage=1
\pageno=1

\jnlstyle

\jl{1}

\overfullrule=0pt

\title{Radial Fredholm perturbation in the two-dimensional Ising
model and gap-exponent relation}[Radial Fredholm perturbation]

\author{Dragi Karevski\dag, Lo\"\ii c Turban\dag\ and  Ferenc
Igl\'oi\ddag}[D Karevski \etal]

\address{\dag Laboratoire de Physique du Solide\footnote{\S}{Unit\'e de
Recherche associ\'ee au CNRS no 155}, Universit\'e Henri Poincar\'e
(Nancy 1), BP 239,\hfill\break  F-54506~Vand\oe uvre l\'es Nancy Cedex,
France}

\address{\ddag Reasearch Institute for Solid State Physics, P.O. Box
49,\hfill\break H-1525~Budapest 114, Hungary\hfill\break and\hfill\break
Institute for Theoretical Physics, Szeged University, Aradi V. tere
1,\hfill\break H-6720~Szeged, Hungary}


\abs
We consider concentric circular defects in the
two-dimensional Ising model, which are distributed according to a
generalized Fredholm sequence, i. e. at exponentially increasing radii.
This type of aperiodicity does not change the bulk critical behaviour but
introduces a marginal extended perturbation. The critical exponent of the
local magnetization is obtained through finite-size scaling, using a
corner transfer matrix approach in the extreme anisotropic limit. It
varies continuously with the amplitude of the modulation and is closely
related to the magnetic exponent of the radial Hilhorst-van Leeuwen
model. Through a conformal mapping of the system onto a strip, the
gap-exponent relation is shown to remain valid for such an aperiodic
defect.

\endabs

\pacs{05.50.+q, 64.60.Cn, 64.60.Fr}
\submitted
\date

\section{Introduction}
The influence of a layered aperiodic modulation of the couplings on the
critical behaviour of the two-dimensional Ising model has been much
studied recently~(see~\cite{luck93} and references therein).
Such a  perturbation may be relevant, marginal or irrelevant, depending
on the sign of the crossover exponent~\cite{luck93,igloi93}
$$
\Phi=1+\nu(\omega-1)\; ,
\label{e1.1}
$$
which involves the bulk correlation length exponent~$\nu$ and the
wandering exponent of the aperiodic
sequence~$\omega$~\cite{queffelec87,dumont90}.

In the case of a marginal perturbation, $\Phi\!=\!0$, some exact results
have been obtained for the Ising model. Critical exponents are found to
vary continuously with the modulation
amplitude~[1,2,5--8] and, when
the critical coupling is shifted, the bulk critical point becomes
strongly anisotropic, i. e. the exponents of the correlation length,
parallel and perpendicular to the layers, are different and their ratio
varies continuously with the amplitude of the modulation~\cite{berche95}.

For some aperiodic sequences, the defect density vanishes in the
thermodynamic limit, there is no shift in the critical coupling and the
bulk critical properties remain unchanged. The  Fredholm
sequence~\cite{dekking83}, which belongs to this class and leads to a
marginal perturbation for the layered Ising model, has been recently
studied on a semi-infinite system~\cite{karevski95}. Continuously varying
surface exponents were obtained and it was shown that this type of
aperiodic perturbation, which happens to be very regular, can be
considered as a discrete realisation of the Hilhorst-van~Leeuwen (HvL)
model~\cite{hilhorst81}, i. e. a semi-infinite Ising model with smoothly
varying couplings.

In the present paper, we study the critical properties of a radial
aperiodic perturbation in the two-dimensional Ising model. Instead of the
parallel line defects of the layered system, we consider concentric
circular defects distributed according to the Fredholm sequence. This
type of perturbation is closely related to the radial HvL defect~[11--13].

Our main motivation is to check the validity of the gap-exponent
relation~\cite{pichard81,luck82} which follows from the transformation
of the critical correlation functions under the conformal mapping of the
original system onto a strip~\cite{cardy84}. Since conformal
transformations cannot be used on strongly anisotropic
systems~\cite{henkel94}, such a relation is excluded in the case of bulk
aperiodic perturbations. But it is known to be satisfied in some
marginally inhomogeneous systems at the critical point, provided the
inhomogeneity is properly transformed, as shown for a narrow line
defect~\cite{turban85,henkel89} and later for extended defects of the HvL
type~\cite{burkhardt90,igloi90,bariev91,turban91}.

The structure of the paper is the following. In section 2, we present the
model and recall the properties of the Fredholm sequence. In section 3,
the local magnetization is studied by the corner transfer matrix method
of Peschel and Wunderling~\cite{peschel92} and the local critical
exponent is compared to the value for the HvL defect. In section 4, the
validity of the gap-exponent relation is discussed and our conclusions
are given in section 5.

\section{The radial Fredholm perturbation}
We consider, in the $(\rho,\vartheta)$-plane, a two-dimensional Ising
model with Hamiltonian
$$
-\beta H=-\beta H_c+g\sum_{p=-\infty}^{+\infty}\int\varepsilon(\rho,
\vartheta)\ \delta(\rho-\rho_p)\,\rho\d\rho\d\vartheta\;
,\qquad\rho_p=m^p\; , \label{e2.1}
$$
in the continuum limit. Here $H_c$ is the critical Hamiltion of
the unperturbed system, $\varepsilon$ is  the energy density operator and
the energy perturbation, with an amplitude $g$ per unit length, is
located on circles with radii $\rho_p\!=\! m^p$.

The contribution of negative values of $p$   to the perturbation is
irrelevant since it renormalizes to a point defect with a finite
amplitude. The local critical behaviour, which is governed by the long
distance behaviour of the perturbation, remains unaffected if one
considers circles with radii distributed according to the generalized
Fredholm sequence~\cite{karevski95}, i. e. with $\rho_p\!=\! m^p\!+\!1$,
$p\!\geq\!0$ and integer values of $m\!>\!1$ .

This aperiodic sequence, which is the characteristic sequence of the
powers of $m$, follows from the substitution on the three letters $A$, $B$ and
$C$:
$$
\eqalign{
A\to {\cal S}(A)&=A\  B\  C\ C\  \cdots\  C\cr
B\to {\cal S}(B)&=B\  C\  C\ C\  \cdots\  C\cr
C\to {\cal S}(C)&=\underbrace{C\  C\  C\ C\  \cdots\  C}_{m}\cr}
\label{e2.2}
$$
With words of length $m\!=\!2$, one recovers the usual Fredholm
substitution~\cite{dekking83}.
Starting the substitution with $A$ and associating to the $k$-th  letter
in the sequence the digit $f_k\!=\!0$ for $A$ and $C$, and $f_k\!=\!1$
for $B$, after $n$ iterations and with $m\!=\!2$, one obtains:
$$
\fl\matrix{
n=0\quad&A&&&&&&&&&&&&&&&\cr
n=1\quad&A&B&&&&&&&&&&&&&&\cr
n=2\quad&A&B&B&C&&&&&&&&&&&&\cr
n=3\quad&A&B&B&C&B&C&C&C&&&&&&&&\cr
n=4\quad&A&B&B&C&B&C&C&C&B&C&C&C&C&C&C&C\cr
f_k\quad&0&1&1&0&1&0&0&0&1&0&0&0&0&0&0&0\cr}
\label{e2.3}
$$
which gives $f_k\!=\!1$ when $k\!=\!2^p+1$.

The substitution matrix has eigenvalues $m$, 1, 1~\cite{karevski95}, so
that the wandering exponent $\omega$ vanishes, leading to a marginal
layered perturbation according to~\ref{e1.1}. It is easy to verify that
the perturbation is also marginal with a radial defect. The total
perturbation inside a circle with radius $R\gg1$ is obtained by summing
over the contributions of the circles up to $p_{max}\approx\ln R/\ln m$,
$$
2\pi g\sum_{p=0}^{p_{max}}m^p\approx{2\pi mgR\over m-1}
\label{e2.4}
$$
and the average perturbation per unit surface at a length scale $R$ is
given by
$$
\overline{g}(R)\sim{g\over R}\; .
\label{e2.5}
$$
Under rescaling by a factor $b=R/R'$, this thermal perturbation transforms
according to
$$
{g'\over R'}=b^{1/\nu}{g\over R}
\label{e2.6}
$$
so that, with $\nu\!=\!1$ in the two-dimensional Ising model, $g$ is
scale invariant, i. e. the radial aperiodic perturbation is
marginal. Furthermore, according to~\ref{e2.5}, the average perturbation
density vanishes in the thermodynamic limit, leaving the bulk critical
point unchanged.

\section{Corner transfer matrix study of the local magnetization}

In this section, the critical behaviour of the local magnetization at
the center of the defect will be deduced from a finite-size scaling
analysis at the bulk critical point. This can be achieved using the
corner transfer matrix method of Peschel and Wunderling~\cite{peschel92}
which allows a study of rotation symmetric defects,  while working on a
lattice.

{\par\begingroup\parindent=0pt\medskip
\epsfxsize=9truecm
\topinsert
\centerline{\hglue22truemm \epsfbox{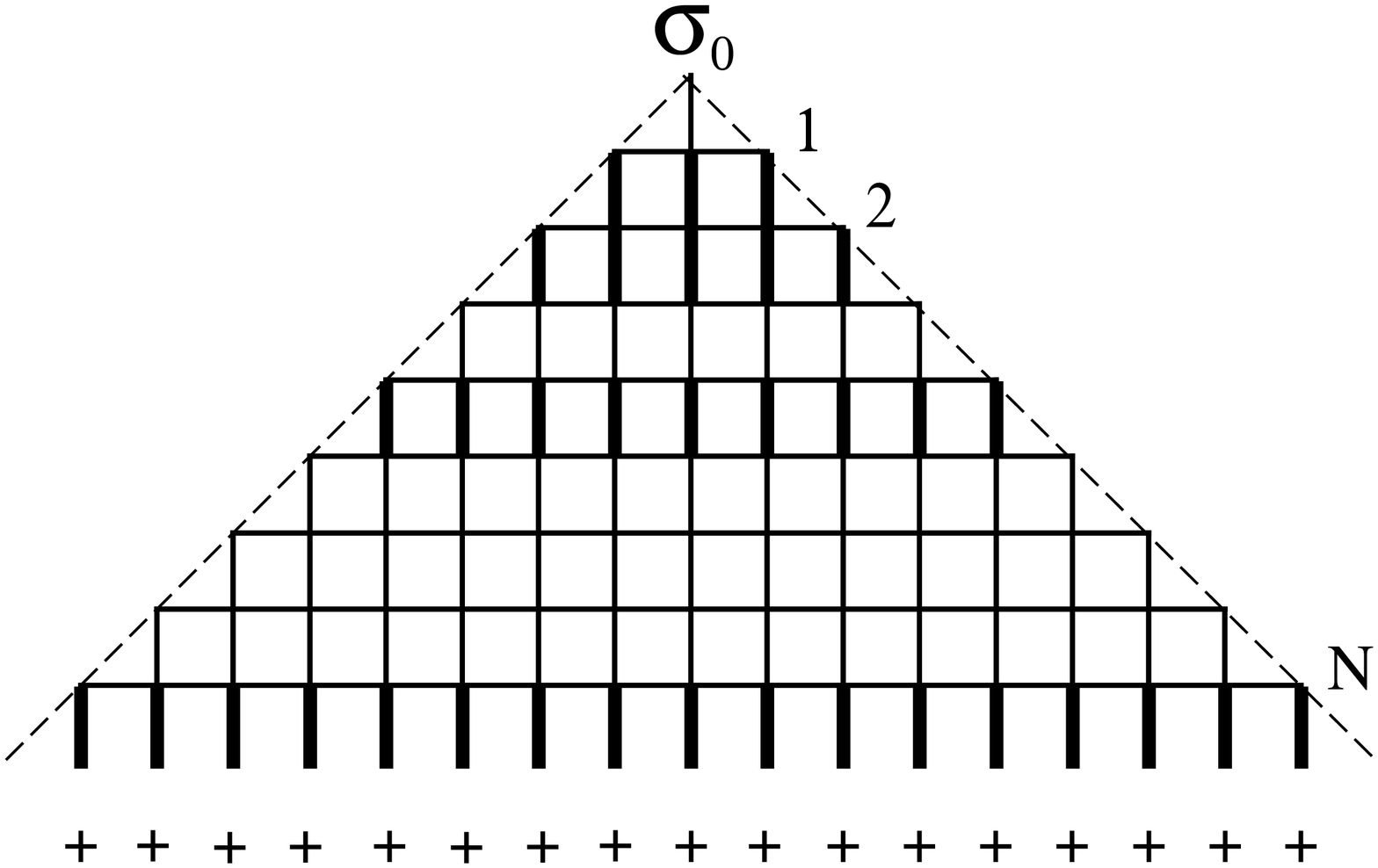}}
\smallskip
\figure{Sector of the anisotropic Ising square lattice which is used to
construct the corner transfer matrix. The horizontal couplings are
constant and equal to $K_1$. The vertical couplings follow the Fredholm
sequence and are equal either to $K_2$ or to $rK_2$ (heavy lines). The
Ising spins on the last row are held fixed in order to calculate the
local magnetization.} \endinsert
\endgroup
\par}

{\par\begingroup\parindent=0pt\medskip
\epsfxsize=9truecm
\midinsert
\centerline{\epsfbox{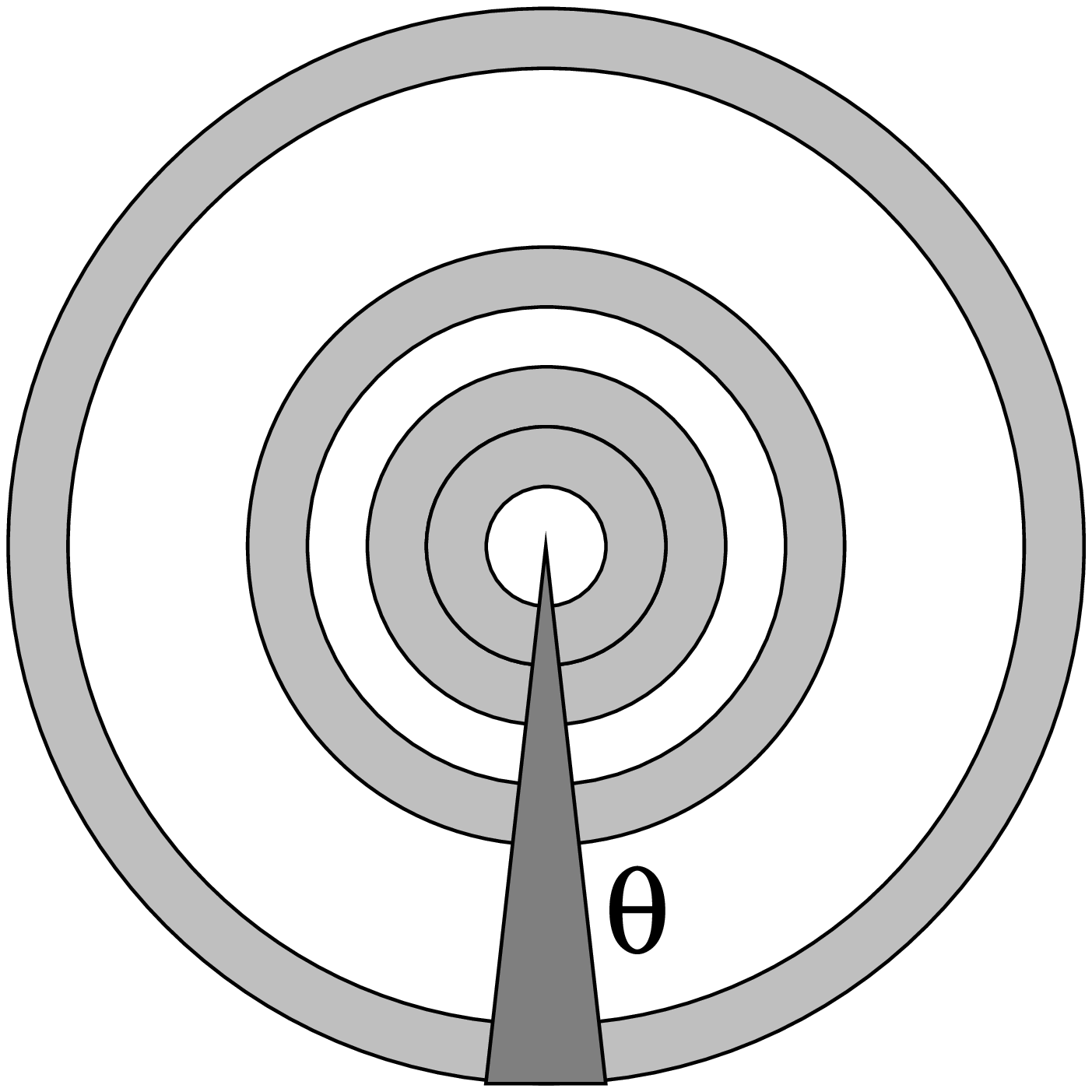}}
\smallskip
\figure{Through a rescaling of the lattice parameters, one restores
isotropy near the critical point. The sector of figure~1 (dark triangle)
has now a vanishing opening angle $\theta$. Covering the plane with such
sectors, one generate the radial Fredholm defect (grey circles).}
\endinsert
\endgroup
\par}

We first consider the sector of a lattice Ising model shown in figure~1.
With the same lattice parameter $a_1\!=\! a_2\!=\! a$ in both directions,
the opening angle is $\theta\!=\pi/2$. There are $N+1$ horizontal rows
with fixed boundary conditions on the last one. The interactions are
between first-neighbour spins, constant and equal to $K_1$ in the
horizontal direction, while they are given by $K_2(k)$ and depend on the
row index, $k=0,N$, in the vertical direction. This dependence follows
the generalized Fredholm sequence, so that the vertical couplings take
the form    $$
K_2(k)=K_2\, r^{f_{k+1}}\; ,
\label{e3.1}
$$
where $r$ is the modulation factor and the $f_k$s are the digits defined
in equation~\ref{e2.3}.

Let us take the extreme anisotropic limit, $K_2\to0$ and $K_1\to\infty$,
while keeping the ratio $\lambda\!=\! K_2/K_1^*$ fixed
($K_1^*\!=\!-\case{1}{2}\ln\tanh K_1\to0$ is a dual coupling). On the
square lattice, the correlation lengths become different in the two
directions with $\xi_2/\xi_1\simeq2K_1^*\ll1$ near the critical point. In
order to recover an isotropic system which properly describes the
continuum problem of the last section, one has to rescale the lattice
parameters such that $\xi_1a_1\!=\!\xi_2a_2$, which gives
$a_1\!=\!2K_1^*\to0$  if one takes $a_2$ for the unit lentgth. The
opening angle of the sector is now reduced to $\theta\!=\!4K_1^*$ and the
number of sectors needed to cover the plane, $$
n={\pi\over 2K_1^*}\; ,
\label{e3.2}
$$
goes to infinity in the extreme anisotropic limit.

In this way, as shown in figure~2, the rotation symmetry of  the original
problem is restored and the system becomes continuous along the defects.
The perturbation per unit length is $(r-1)K_2/a_1$ which allows us to
identify the continuum parameter $$
g=\case{1}{2}(r-1)
\label{e3.3}
$$
at the critical point $\lambda_c=1$.

In the extreme anisotropic limit, the Baxter corner transfer matrix ${\cal
T}$~\cite{baxter82} takes the simple form
$$
{\cal T}=\exp\left(-\case{1}{2}\theta{\cal H}\right)
\label{e3.4}
$$
where $\theta$ is the infinitesimal opening angle of the sector and
${\cal H}$ is the Hamiltonian of the inhomogeneous quantum Ising chain:
$$
\fl{\cal H}=-\case{1}{2}\sum_{k=1}^{N-1}2k\; \sigma_k^z-
\case{1}{2}\sum_{k=0}^{N-1}(2k+1)\;\lambda(k)\; \sigma_k^x\sigma_{k+1}^x\; ,
\qquad\lambda(k)=\lambda r^{f_{k+1}}\; .\label{e3.5}
$$
The $\sigma$s are Pauli spin operators and the coupling $\lambda(k)$ has
the same aperiodic modulation as $K_2$ in equation~\ref{e3.1}.

The magnetization of the central spin, given by
$$
m_0={{\rm Tr}\, (\sigma_0^x\sigma_N^x{\cal T}^n)\over
{\rm Tr}\, ({\cal T}^n)}\; ,
\label{e3.6}
$$
can be reexpressed in terms of the Fermi operators  which diagonalize the
corner transfer matrix~\cite{jordan28,lieb61}. In this way, one obtains
the local magnetization as a product,
$$
m_0=\prod_{\nu=1}^N\tanh\left(\case{1}{2}\pi\epsilon_\nu\right)\; ,
\label{e3.7}
$$
where the $\epsilon_\nu$s are the $N$ nonvanishing diagonal fermion
excitations of the quantum chain\footnote{$\Vert$}{The lowest excitation
$\epsilon_0$ vanishes due to the fixed boundary conditions.} (see
reference~\cite{peschel92} or appendix B in reference~\cite{igloi93b} for
details).

{\par\begingroup\parindent=0pt\medskip
\epsfxsize=9truecm
\topinsert
\centerline{\epsfbox{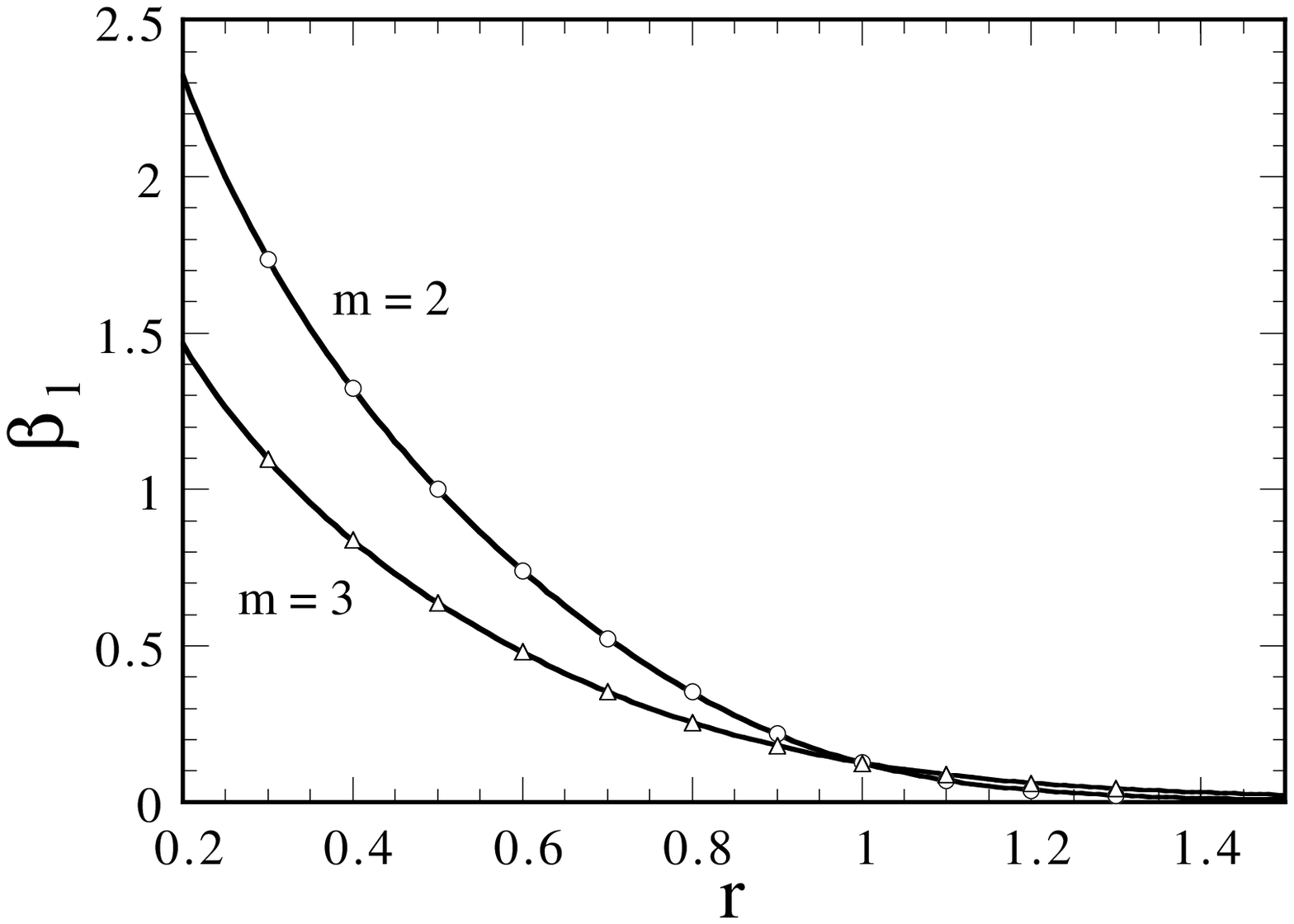}}
\smallskip
\figure{Variation of the local magnetization exponent $\beta_l$ with the
modulation factor $r$ for the generalized Fredholm sequence with $m\!=\!2$
and $m\!=\!3$. The lines correspond to the conjectured expression of
equation~(3.9) and the points are finite-size scaling estimates.}
\endinsert  \endgroup
\par}

\midinsert
\bigskip
\table{Extrapolated finite-size estimates for the local magnetization
exponent $\beta_l$ and the shift $\Delta$ contributed by the localized
mode when $r<1$, as functions of the modulation factor $r$ for the
Fredholm radial defect with $m=2$ and $m=3$. The figures in brackets give
the estimated uncertainty in the last digit. Each second column gives the
expected values of equation~(3.9).}[w] \align\L{#}&&\L{#}\cr \br
&\centre{4}{$m=2$}&\centre{4}{$m=3$}\cr  \ns &\crule{4}&\crule{4}\cr
$r$&\centre{2}{$\beta_l$}&\centre{2}{$\Delta$}
&\centre{2}{$\beta_l$}&\centre{2}{$\Delta$}\cr
\ns
\crule{1}&\crule{2}&\crule{2}&\crule{2}&\crule{2}\cr
0.3&1.7373&1.7370&1.73697(6)&1.73696559&1.0973&1.0963&
1.0959031(5)&1.09590327\cr
0.4&1.3221&1.3220&1.32192811(4)&1.32192809&0.8373&0.8356&
0.834043(1)&0.83404377\cr
0.5&1.0010&1.0006&0.9999999(1)&1&0.6382&0.6362&0.630929(2)&0.63092975\cr
0.6&0.7405&0.7398&0.736964(3)&0.73696559&0.4801&0.4780&0.46493(6)&
0.46497352\cr
0.7&0.5254&0.5245&0.51462(7)&0.51457317&0.3538&0.3522&0.326(2)&
0.32465952\cr
0.8&0.3505&0.3498&0.32191(5)&0.32192809&0.2546&0.2537&0.211(6)&
0.20311401\cr
0.9&0.2168&0.2166&0.1517(3)&0.15200309&0.1795&0.1793&0.101(4)&
0.09590327\cr
1.0&0.1250&0.125&\centre{1}{---}&\centre{1}{---}&0.1250&0.125&
\centre{1}{---}&\centre{1}{---}\cr
1.1&0.0692&0.0690&\centre{1}{---}&\centre{1}{---}&0.0870&0.0868&
\centre{1}{---}&\centre{1}{---}\cr
1.2&0.0382&0.0377&\centre{1}{---}&\centre{1}{---}&0.0611&0.0605&
\centre{1}{---}&\centre{1}{---}\cr
1.3&0.0215&0.0208&\centre{1}{---}&\centre{1}{---}&0.0436&0.0425&
\centre{1}{---}&\centre{1}{---}\cr \br \endalign \endtable \endinsert

The excitations of the Hamiltonian~\ref{e3.5} at the critical point
$\lambda_c=1$ have been obtained numerically on chains with $N=m^p+1$
spins with $p=4$ to $16$ for $m=2$ and $p=3$ to $9$  for $m=3$. The
modulation factor $r$ has been varied from~$0.3$ to $1.3$ with steps of
$0.1$. The critical value of  $m_0$ on a finite system vanishes as
$N^{-x_l}$, where the scaling dimension of the local magnetization, $x_l$,
is continuously varying in the present case. This exponent is also equal
to the local magnetization exponent~$\beta_l$ since $\nu=1$ in the $2d$
Ising model. The finite-size estimates  for $\beta_l$ obtained from
sequence extrapolations using the BST algorithm~\cite{henkel88} are shown
in figure~3.

The layered Fredholm modulation on a semi-infinite system is known to
lead to the same critical behaviour as the HvL model with couplings
$K_2(k)=K_2(1+\alpha/k)$ if one makes the
correspondence~\cite{karevski95}
$$
\alpha\to{\ln r\over\ln m}\; .
\label{e3.8}
$$
Assuming the same relation for the radial defect and using the analytical
result of reference~\cite{peschel92} for the HvL model, we are led to the
following conjecture for the Fredholm exponent\footnote{\P}{The amplitude
$\alpha$ is half the one used in reference~\cite{peschel92}.}
$$
\fl\beta_l=2\left({\ln r\over\ln m}\right)^2\int_0^\infty\d
u\;{\sinh^2u\over\sinh\left(2\pi\left\vert{\ln r\over\ln
m}\right\vert\cosh\; u\right)}\quad\left(+\left\vert{\ln r\over\ln
m}\right\vert\right)\; . \label{e3.9}
$$
Here the quantity in brackets is a shift $\Delta$, contributed by a
localized mode, which has to be added when $r<1$. Our numerical estimates
for $\beta_l$ are in reasonable agreement with this expression as shown in
table~1. The uncertainties on $\beta_l$, estimated by comparing different
extrapolated values given by the BST algorithm, are much smaller than the
actual deviations from the conjectured values. Such a behaviour is known
to occur when logarithmic corrections to scaling are
present~\cite{henkel88}. The correspondence~\ref{e3.8} is strongly
supported by the extrapolated values of the shift $\Delta$ for which the
agreement with the expected values is excellent.

\section{Conformal mapping and gap-exponent relation}

Let us now consider the transformation of the marginal perturbation
in the continuum limit, equation~\ref{e2.1}, under the logarithmic
conformal mapping~\cite{cardy84}
$$
w={L\over2\pi}\ln z\; ,\qquad w=u+\i v\; ,\qquad
z=\rho\e^{\i\vartheta}\; ,
\label{e4.1}
$$
which transforms the whole plane into a strip
($-\!\infty\!<\!u\!<\!+\!\infty$, $0\!<\!v\!<\! L$) with periodic boundary
conditions in the $v$-direction.  The local dilatation factor is
$b(z)\!=\!\vert\d w/\d z\vert^{-1}\!=\!2\pi\rho/L$ and the amplitude of
the thermal perturbation,  $t(z)\!=\! g\sum_p\delta(\rho-\rho_p)$, is
changed into:  $$\fl\eqalign{
t(w)&=b(z)^{1/\nu}t(z)=g\left({2\pi\rho\over
L}\right)^{1/\nu}\sum_{p=-\infty}^{+\infty}\delta\left(\exp{2\pi u\over
L}-m^p\right)\cr
&=g\left({2\pi\over
L}\right)^{1/\nu}\sum_{p=-\infty}^{+\infty}\delta\left\{\exp\left[{2\pi
u\over L}\left(1-{1\over\nu}\right)\right]-\exp\left[p\,\ln m-{2\pi
u\over\nu L}\right]\right\}}
\label{e4.2}
$$
With the Ising model, $\nu\!=\!1$, the perturbation is marginal, and the
first part of the $u$-dependence is eliminated so that:
$$\eqalign{
t(w)&=g\left({2\pi\over L}\right)\sum_{p=-\infty}^{+\infty}
\delta\left\{1-\exp\left[-{2\pi\over L}\left(u-p\,{L\ln
m\over2\pi}\right)\right]\right\}\cr
&=g\sum_{p=-\infty}^{+\infty}
\delta\left(u-p\,{L\ln m\over2\pi}\right)}
\label{e4.3}
$$
As shown in figure~4, the perturbation now consists of straight line
defects, the distance between two successive lines,
$$
L_m\!=\! L\ln m/2\pi\; ,
\label{e4.3b}
$$
being proportional to the width of the strip $L$. The discrete dilatation
invariance of the radial defect on the plane has been transformed into a
discrete translation invariance along the strip.

{\par\begingroup\parindent=0pt\medskip
\epsfxsize=9truecm
\topinsert
\centerline{\epsfbox{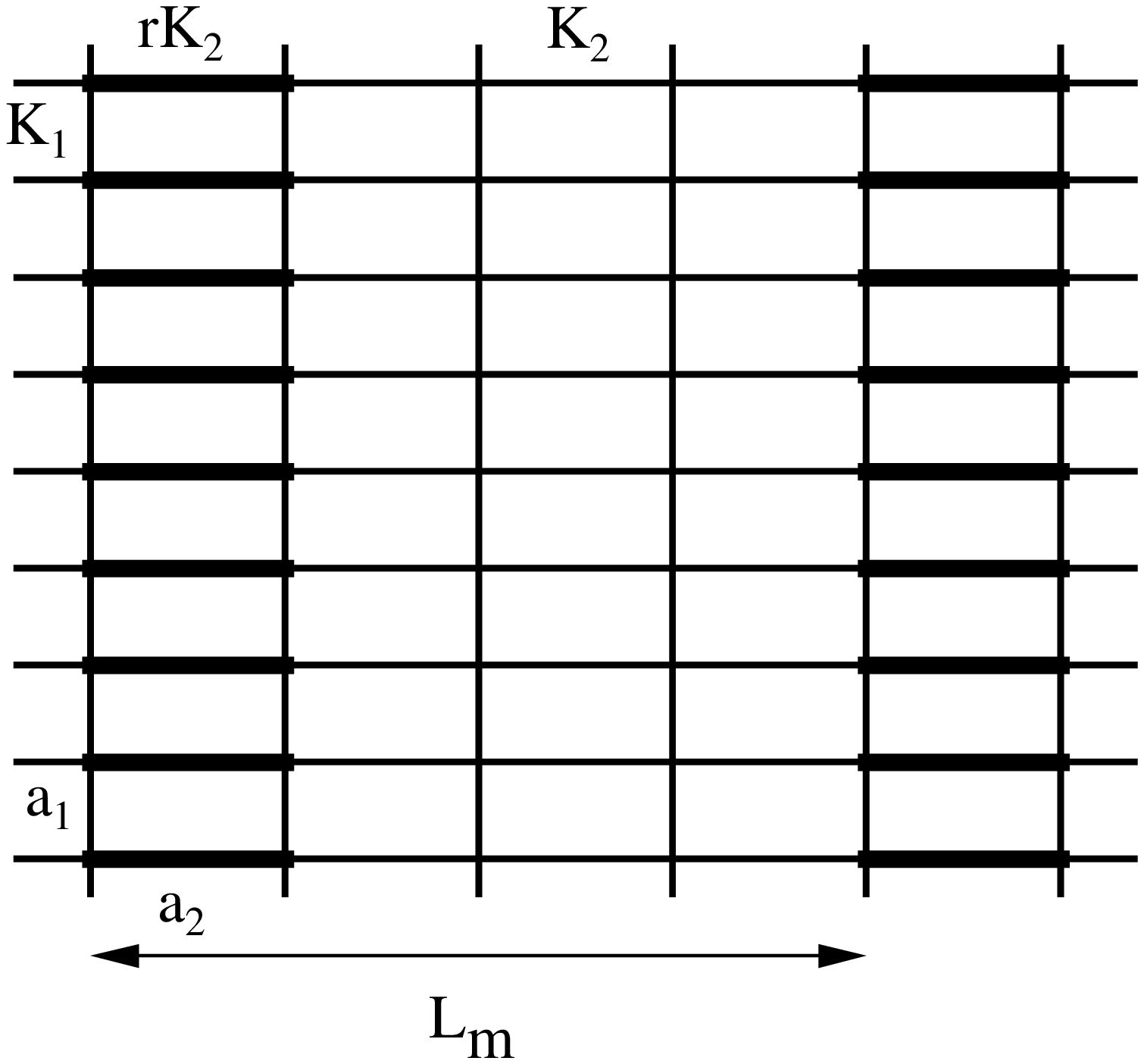}}
\smallskip
\figure{The logarithmic conformal mapping transforms the radial aperiodic
Fredholm defect in the plane into a periodic defect on the strip. The
period $L_m$ is proportional to the strip width $L$.}
\endinsert
\endgroup
\par}

In the case of the HvL radial defect, the continuous dilatation
invariance of the perturbation, $t(z)\!=\! g/\rho$ on the plane, leads to
a constant deviation from the critical coupling $t(w)\!=\! g(2\pi/L)$ on
the strip~\cite{bariev91,turban91,igloi93b}. If $\xi_\phi$ denotes the
correlation length associated with the spin-spin or the energy-energy
correlations on the off-critical isotropic strip, it satisfies the
following finite-size scaling relation~\cite{igloi93b}
$$
\xi_\phi^{-1}(t,L)=L^{-1}X_\phi(cL^{1/\nu}t)\; ,
\label{e4.4}
$$
where the gap scaling function $X_\phi(\tau)$ is
universal~\cite{privman84}, $t$ is the deviation from the bulk critical
temperature and $c$ is a nonuniversal constant. The defect scaling
dimension follows from the gap-exponent relation~[14--16] with:
$$
x_l^\phi={L\over2\pi}\xi_\phi^{-1}(t,L)={1\over2\pi}X_\phi(cL^{1/\nu}t)\; .
\label{e4.5}
$$
It is continuously varying with the defect amplitude $g$, as expected for
a marginal perturbation, but it varies in a universal way: different
models belonging to the same class of universality will show the same
dependence.

With the Fredholm radial defect, which gives  a periodic layered
perturbation in the strip geometry, one could directly deduce the gap
(inverse correlation length) from the appropriate transfer matrix on the
strip and use~\ref{e4.5}.  Alternatively, one may estimate the deviation
$t$ from the bulk critical temperature in~\ref{e4.5} which is associated
with the periodic defect. A measure of this deviation is provided by the
shift in the critical temperature induced by the same perturbation (i. e.
parrallel line defects with a fixed distance $L_m$) on the infinite plane.

In order to avoid the calculation  of the nonuniversal constant $c$, we
shall proceed via a comparison to the HvL problem. We use the same extreme
anisotropic limit as for the corner transfer matrix, with vertical
couplings $K_1\!\to\!\infty$, horizontal couplings $K_2\!\to\!0$ and
vertical ladder defects  corresponding to modified couplings
$$
K'_2\!=\! rK_2
\label{e4.6}
$$
as shown in figure~4. As above, a length rescaling ($a_2\!=\!1$,
$a_1\!=\!2K_1^*$) is understood in order to restore isotropy.  For the
radial HvL problem the perturbed interactions $K_2(k)\!=\!
K_2(1+\alpha/k)$ lead to an homogeneous shift of the horizontal couplings
with    $$
K'_2=K_2\left(1+\alpha{2\pi\over L}\right)\; .
\label{e4.7}
$$

The critical temperature of a periodic layered anisotropic system
follows from the relation~\cite{wolff81}
$$
\sum_jK_2^*(j)=\sum_jK_1(j)\; ,
\label{e4.8}
$$
where $K_2$-bonds are perpendicular to the layers and the sums are over a
period. In the case of the Fredholm problem, this leads to
$$
(L_m-1)K_2^*+{K'_2}^*=L_mK_1\; ,
\label{e4.9}
$$
where, in the extreme anisotropic limit, equation~\ref{e4.6} leads to
$$
{K'_2}^*=K_2^*-\case{1}{2}\ln r\; .
\label{e4.10}
$$
Putting \ref{e4.10} into \ref{e4.9} gives the criticality condition:
$$
K_2^*-{\ln r\over2L_m}=K_1\; .
\label{e4.11}
$$
The corresponding relation for the HvL problem is obtained
using~\ref{e4.9} with $L_m\!=\!1$. Since the modified coupling
in~\ref{e4.7} corresponds to $r\!=\!1+\alpha(2\pi/L)$, with $L\gg1$
equation~\ref{e4.10} gives: $$
{K'_2}^*\approx K_2^*-\alpha{\pi\over L}
\label{e4.12}
$$
and finally the criticality condition reads:
$$
K_2^*-\alpha{\pi\over L}=K_1\; .
\label{e4.13}
$$
Comparing equations~\ref{e4.11} and~\ref{e4.13}  and using~\ref{e4.3b}
leads to the conjectured correspondence of equation~\ref{e3.8}.

Besides the local magnetization exponent of equation~\ref{e3.9}, the
gap-exponent relation also gives the local energy exponent of the radial
Fredholm defect as~\cite{turban91}
$$
x_l^e=1+2\left({\ln r\over\ln m}\right)^2+O\left[\left({\ln r\over\ln
m}\right)^4\right] \label{e4.14}
$$
where, due to the self-duality of the Ising model, only even
powers of ${\ln r/\ln m}$ enter the expansion~\cite{bariev92}.

\section{Conclusion}

The close relationship between the discrete aperiodic Fredholm defect and
the continuous HvL inhomogeneity in the two-dimensional Ising model has
been established in a study of the surface critical behaviour
of the layered system in~\cite{karevski95}.  It has been also recently
verified for the bulk layered Fredholm defect~\cite{karevski95b}.

In the present paper the validity of this  connection --which is
summarized in equation~\ref{e3.8}-- is further extended to radially
symmetric defects. The inhomogeneity in this case corresponds to an
infinite sequence of concentric circles with exponentially increasing
radii. The separation of distant circles becoming very large, the density
of defect bonds goes to zero, thus the bulk critical behaviour of the
system remains unchanged. However an infinite sequence of defect circles
modifies the local critical behaviour at the centre of the defect.

Our second observation, concerning  the validity of the gap-exponent
relation for the Fredholm defect,  is somewhat unexpected. It was known
till now~\cite{igloi93} that some aspects of conformal invariance,
including the gap-exponent relation, are still satisfied for some
marginally inhomogeneous systems, which either contain a finite number of
defect lines or display a smooth variation. In the Fredholm problem the
number of defect circles is infinite, furthermore the perturbation changes
the continuous dilatation symmetry into a discrete one.

A similar analysis of the layered Fredholm defect
problem~\cite{karevski95,karevski95b}   reveals that the gap-exponent
relation stays valid in this case, too. Now the transformed perturbation
on the strip is still periodic with period $L_m$ as given in
equation~\ref{e4.3b}, however its shape is much more complicated than for
the radial defect. Taking the anisotropic limit and considering the
product of $L_m$ successive transfer matrices one can establish the same
relation with the HvL model as in~\ref{e3.8}. These observations lead us
to conjecture the validity of the gap-exponent relation for any marginal
perturbation which does not modify the bulk critical behaviour of the
system.

If the marginal perturbation extends over the volume of the system with a
nonvanishing density,  the gap-exponent relation is generally no longer
valid. As mentioned in the introduction,  this type of marginal aperiodic
perturbations lead to strongly anisotropic systems~\cite{berche95} which
cannot be transformed using conformal techniques.

 \ack The Nancy-Budapest collaboration  is supported by the CNRS and the
Hungarian Academy of Sciences. FI acknowledges financial support from the
Hungarian National Research Fund under grant No OTKA TO12830. The
numerical work in Nancy was supported by CNIMAT under project No 15q5C93b.

\numreferences

\bibitem{luck93} {Luck J M 1993} {\it J. Stat. Phys.} {\bf 72} {417}

\bibitem{igloi93} {Igl\'oi F 1993} {\JPA} {\bf 26} {L703}

\bibitem{queffelec87} {Queff\'elec M 1987} {\it Substitution Dynamical
Systems-Spectral Analysis}\ {Lecture Notes in Mathematics vol~1294
ed.~A~Dold and B~Eckmann (Berlin: Springer) p~97}

\bibitem{dumont90} {Dumont J M 1990} {\it Number Theory and Physics}\
{Springer Proc. Phys. vol~47 ed.~J~M~Luck, P~Moussa and  M~Waldschmidt
(Berlin: Springer) p~185}

\bibitem{turban94a} {Turban L, Igl\'oi F and Berche B 1994} {\PR\ {\rm
B}}\ {\bf 49} {12695}

\bibitem{turban94b}{Turban L, Berche P E and Berche B 1994} {\JPA}\ {\bf
27} {6349}

\bibitem{karevski95} {Karevski D, Pal\'agyi G and Turban L 1995} {\JPA}\
{\bf 28} {45}

\bibitem{berche95} {Berche B, Berche P E, Henkel M, Igl\'oi F, Lajk\'o P,
Morgan S and Turban L 1995} {\JPA}\ {\bf 28} {L165}

\bibitem{dekking83} {Dekking M, Mend\`es-France M and van der
Poorten A 1983} {\it Math. Intelligencer}\ {\bf 4} {130}

\bibitem{hilhorst81} {Hilhorst H J and van Leeuwen J M J 1981} {\PRL}\
{\bf 47} {1188}

\bibitem{bariev91} {Bariev R Z and Peschel I 1991} {\JPA}\ {\bf 24} {L87}

\bibitem{turban91} {Turban L 1991} {\PR\  {\rm B}}\ {\bf 44} {7051}

\bibitem{peschel92} {Peschel I and Wunderling R 1992} {\it Ann. Physik}\
{\bf 1} {125}

\bibitem{pichard81} {Pichard  J L and Sarma G 1981} {\JPC}\ {\bf 14} {L127}

\bibitem{luck82} {Luck J M 1982} {\JPA}\ {\bf 15} {L169}

\bibitem{cardy84} {Cardy J L 1984} {\JPA}\ {\bf 17} {L385}

\bibitem{henkel94} {Henkel M 1994} {\it J. Stat. Phys.}\ {\bf 75} {1023}

\bibitem{turban85} {Turban L 1985} {\JPA}\ {\bf 18} {L325}

\bibitem{henkel89} {Henkel M, Patk\'os A and Schlottmann M 1989} {\NP\
{\rm B}}\  {\bf 314} {609}

\bibitem{burkhardt90} {Burkhardt T W and Igl\'oi F 1990} {\JPA}\  {\bf
23} {L633}

\bibitem{igloi90} {Igl\'oi F, Berche B and Turban L 1990} {\PRL}\ {\bf 65}
{1773}

\bibitem{baxter82} {Baxter R J 1982} {\it Exactly Solved Models in
Statistical Mechanics} {(London: Academic Press)} {p~363}

\bibitem{jordan28} {Jordan P and Wigner E 1928} {\ZP}\ {\bf 47} {631}

\bibitem{lieb61} {Lieb E H, Schultz T D and Mattis D C 1961} {\APNY}\
{\bf 16} {406}

\bibitem{igloi93b} {Igl\'oi F, Peschel I and Turban L 1993} {\it Adv.
Phys.}\  {\bf 42} {683}

\bibitem{henkel88} {Henkel M and Sch\"utz G 1988} {\JPA}\  {\bf 21} {2617}

\bibitem{privman84} {Privman V and Fisher M E 1984} {\PR\ {\rm B}}\ {\bf
30} {322}

\bibitem{wolff81} {Wolff W F, Hoever P and Zittartz J 1981} {\ZP\ {\rm
B}}\ {\bf 42} {259}

\bibitem{bariev92} {Bariev R Z and Turban L 1992} {\PR\ {\rm B}}\ {\bf
45} {10761}

\bibitem{karevski95b} {Karevski D 1995} {}{}{(unpublished)}

\vfill\eject
\bye